\documentclass[aps,prd,eqsecnum,notitlepage,
superscriptaddress,
nofootinbib,longbibliography]{revtex4-2}

\usepackage{physics}
\usepackage{amsmath,amssymb,amsfonts}
\usepackage{bm}
\usepackage{graphicx}
\usepackage{natbib}
\usepackage[caption=false]{subfig}
\usepackage{hyperref}
\usepackage{array}
\usepackage{float}
\usepackage{xcolor}
\global\arraycolsep=2pt
\DeclareMathOperator{\sgn}{sgn}

\usepackage{tikz,xcolor,hyperref}
\definecolor{lime}{HTML}{A6CE39}
\DeclareRobustCommand{\orcidicon}{%
	\begin{tikzpicture}
	\draw[lime, fill=lime] (0,0) 
	circle [radius=0.16] 
	node[white] {{\fontfamily{qag}\selectfont \tiny ID}};	\draw[white, fill=white] (-0.0625,0.095) 
	circle [radius=0.007];	\end{tikzpicture}
	\hspace{-2mm}}
\foreach \x in {A, ..., Z}{%
	\expandafter\xdef\csname orcid\x\endcsname{\noexpand\href{https://orcid.org/\csname orcidauthor\x\endcsname}{\noexpand\orcidicon}}
	}

\newcommand{\be}{\begin{equation}} 
\newcommand{\ee}{\end{equation}}

\begin{document}

%title%
\title{Quantum friction in the presence of a perfectly conducting plate}

%author information%
%\author{Xin Guo}
\author{Xin Guo\orcidA{}}
\email{guoxinmike@ou.edu}
%\homepage[]{Your web page}
%\thanks{}
%\altaffiliation{}
\affiliation{H. L. Dodge Department of Physics and Astronomy, University of Oklahoma, Norman, Oklahoma 73019, USA}

%\author{Kimball A. Milton}
\author{Kimball A. Milton\orcidB{}}
\email{kmilton@ou.edu}
%\homepage[]{Your web page}
%\thanks{}
%\altaffiliation{}
\affiliation{H. L. Dodge Department of Physics and Astronomy, University of Oklahoma, Norman, Oklahoma 73019, USA}

%%\author{Gerard Kennedy}
\author{Gerard Kennedy\orcidC{}}
\email{g.kennedy@soton.ac.uk}
%%\homepage[]{Your web page}
%%\thanks{}
%%\altaffiliation{}
\affiliation{School of Mathematical Sciences, University of Southampton, Southampton SO17 1BJ, United Kingdom}

%\author{William P. McNulty\orcidD{}}
%%\author{William P. McNulty}
%\email{william.p.mcnulty-1@ou.edu}
%%\homepage[]{Your web page}
%%\thanks{}
%%\altaffiliation{}
%\affiliation{H. L. Dodge Department of Physics and Astronomy, University of Oklahoma, Norman, Oklahoma 73019, USA}

\author{Nima Pourtolami\orcidF{}}
%%\author{Nima Pourtolami}
\email{nima.pourtolami@gmail.com}
%%\homepage[]{Your web page}
%%\thanks{}
%%\altaffiliation{}
\affiliation{National Bank of Canada, Montreal, Quebec H3B 4S9, Canada}

%\author{Yang Li\orcidE{}}
%\email{leon@ncu.edu.cn}
%%\homepage[]{Your web page}
%%\thanks{}
%%\altaffiliation{}
%\affiliation{Department of Physics, Nanchang University, Nanchang 330031, China}

%date%
\date{\today}

%abstract%
\begin{abstract}
A neutral but polarizable particle at rest near a perfectly conducting plate feels a force normal to the surface of the plate, which tends to pull the particle towards the plate. This is the well-known Casimir-Polder force, which has long been theoretically proposed and experimentally observed. In this paper, we explore the transverse frictional force on an atom moving uniformly parallel to a perfectly conducting plate. Although many theoretical predictions can be found for the quantum friction on a particle moving above an imperfect surface, the extreme situation with a perfectly conducting plate seems to have been largely ignored by the theoretical community.  We investigate this ideal case as a natural extension of our previous works on quantum vacuum friction (blackbody friction), and conclude that there does exist a quantum frictional force on an atom moving above a perfectly conducting plate. Very interestingly, the distance dependence, the temperature dependence and even the sign of the frictional force can depend on the polarization state of the atom. For an isotropic atom with a static polarizability, the resultant frictional force is found to be negative definite and therefore remains a true drag. Just above the surface of the plate, the magnitude of the frictional force is twice that of the quantum vacuum friction in the absence of the plate. 

%As expected, the quantum frictional force in the presence of a perfectly conducting plate reduces to the quantum vacuum friction in the limit of $ aT\gg 1 $. Here, $ a $ is the fixed distance between the particle and the plate, while $ T $ is the radiation temperature. Therefore, the new physics is the behavior of the frictional force at short distances or low temperatures. 

%If the particle is only polarizable in the transverse directions, the frictional force is found to be proportional to $ a^{4}T^{12} $ in the limit of $ aT\ll 1 $, and therefore decays to zero as the particle gets very close to the surface. If the particle is only polarizable in the normal direction, we find the frictional force in the limit of $ aT\ll 1 $ precisely quadruples the corresponding quantum vacuum friction, which is independent of $ a $ and proportional to $ T^{8} $. In both of these situations, the frictional force is negative. However, if the particle is polarizable both in the normal direction and in the direction of motion, another positive contribution to the frictional force exists, which goes as $ a^{2} T^{10} $. 

\end{abstract}

%keywords%
%\keywords{}

%\maketitle must follow title, authors, abstract, and keywords%
\maketitle

\section{introduction}
\label{intro}
It is well known that, when a neutral but polarizable particle sits near a perfectly conducting (PC) plate, it feels a force normal to the surface, pulling it towards the plate. This attractive force is often named after Casimir and Polder, who predicted it back in 1948 \cite{Casimir:CP}. And the Casimir-Polder force was first experimentally confirmed by measuring the deflection of a sodium atom beam passing through a gold cavity \cite{Sukenik:CP}. There have been many experiments confirming the existence of the Casimir-Polder force ever since. Another ingenious method, which is suitable for detecting the thermal effects at larger atom-surface separation, is through the measurement of the center-of-mass oscillation frequencies of a rubidium atom Bose-Einstein condensate \cite{Harber:CP, Obrecht:CP}. But, will a force parallel to the surface of the 
PC plate arise when the particle moves parallel to the plate?

Even though the subject of quantum friction (QF) with a dielectric surface has been much discussed in the literature, this more idealized case involving a PC plate seems to have been largely ignored. The lack of discussion of this case may be due to an incorrect intuition arising from the image particle picture. One might think that the interaction between the particle and the PC plate can be entirely mimicked by the particle's interaction with its image. As the particle moves above the plate, the image moves below the plate. Because the plate is perfectly conducting, the image keeps up with the particle and is always located at the mirror position of the particle. Consequently, any interaction between the two would only lie in the direction normal to the surface of the plate and no force in the transverse directions could possibly arise. This reasoning sounds convincing except that it ignores one important aspect: the particle interacts with the blackbody radiation surrounding it even when the plate is taken away. The nonrelativistic discussion of frictional force on particles moving in free space filled with only blackbody radiation can be traced back to the works of Mkrtchian et al. \cite{Mkrtchian:universal} or even Einstein and Hopf \cite{Einstein:Hopf}. Ever since, there has been considerable interest in the subject of blackbody friction/quantum vacuum friction (QVF) \citep{Dedkov:tangential,  Lach:EH, Dedkov:photon, Sinha:dipole}.  Recently, we have also investigated such quantum vacuum frictional effects on a particle moving with relativistic velocities, be it a nondissipative atom \cite{Xin:eqf1} or an intrinsically dissipative nanoparticle \cite{Xin:eqf2}. Now, when a PC plate is added into the configuration, the vacuum field in the vicinity of the plate will be different from that of the free space considered in Refs.~\cite{Xin:eqf1,Xin:eqf2}. We therefore expect the QVF to be modified and become spatially varying in the normal direction.  For convenience of presentation, we will refer to this quantum frictional force on a neutral particle passing above a PC plate as QFPC.

In this paper, we focus on calculating the QFPC for a nondissipative atom. This is somewhat simpler than the  calculation of QFPC for a dissipative nanoparticle, where the temperatures of the particle and of the environment enter the problem independently. The discussion of QFPC for the dissipative nanoparticle is postponed to a subsequent paper.

Throughout the paper, we set $ k_{B}=c=\hbar=1 $ in the analytic expressions. SI units are reinstated in the numerical evaluations.

\section{GENERAL THEORY}
\begin{figure*}
%\begin{tikzpicture}[scale=0.5, cross/.style={path picture={ 
%  \draw[black]
%(path picture bounding box.south east) -- (path picture bounding box.north west) (path picture bounding box.south west) -- (path picture bounding box.north east);
%}}]
%\draw[->](1,5.5)--(2,5.5) node[anchor=west]{$x$};
%\draw[->](1,5.5)--(1,6.5) node[anchor=south]{$z$};
%\node [draw, scale=0.5, circle, cross] at (1,5.5){};
%\node at (0.7,5.2) {$y$};
%\node at (12,5) {$T$};
%\shade[ball color=red] (7,4.5) circle (0.4) node[anchor=south east]{$\bm{\alpha}$};
%\draw[ultra thick, ->] (7.4,4.5) -- (8.4,4.5) node[anchor=south]{$v$};
%\draw[dashed] (7,4.1) to node[right]{$a$} (7,1) ;
%\fill[color=yellow, opacity=0.7] (0,0) rectangle (14,1);
%\node at (12,0.5) {$\epsilon=\infty$};
%\draw[ultra thick, ->] (7.0,4.1) -- (7.0,3.1) node[anchor=east]{$F_{\rm{CP}}$};
%\draw[ultra thick,dotted, ->](6.6,4.5)--(5.6,4.5) node[anchor=east]{$F_{\rm{ric}}$};
%\end{tikzpicture}
\includegraphics[width=0.7\linewidth]{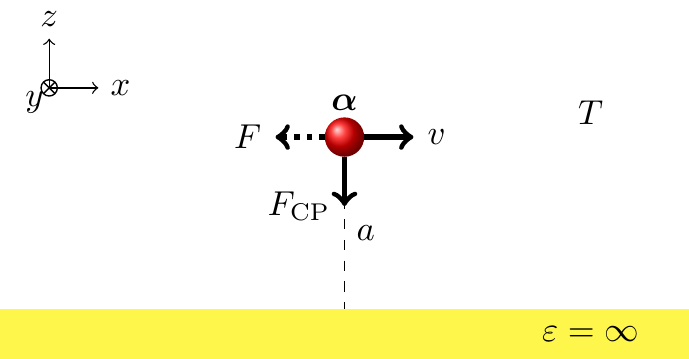}
\caption{Illustration of an atom flying above a PC plate.} 
\label{setup}
\end{figure*}

The physical situation we consider is illustrated in Fig.~\ref{setup}. A PC plate lies in the $ x $-$ y $ plane. An atom is at a distance $ a $ from the plate and moves in the $ x $ direction with constant velocity $ v $. The polarizability of the atom is $ \bm{\alpha}(\omega) $, which could be dispersive in frequency and have different components corresponding to different polarization states of the atom. Since the atom we consider is intrinsically nondissipative, $ \bm{\alpha}(\omega) $ is a real quantity. The radiation background is at finite temperature $ T $. We assume the PC plate is in thermal equilibrium with the radiation background. Because of its motion, the atom is not in equilibrium with the radiation. However, it is guaranteed, by the optical theorem \cite{Berman:OptTrm}, to be in the nonequilibrium steady state (NESS) \cite{Hoshino:NESS}, and it does not have an independent temperature \cite{Xin:eqf1}.
%setup picture

The friction on an atom moving above a general planar surface has actually been calculated in Ref.~\cite{Xin:eqf1} and tabulated for different polarization states in Appendix D therein. We find the formula for quantum friction can be recast into the following more convenient form:
%starting equation
\begin{equation}\label{starting}
F=\int\frac{d\omega}{2\pi}\frac{d^{2}\bm{k}_{\perp}}{(2\pi)^2}\frac{d^{2}\bar{\bm{k}}_{\perp}}{(2\pi)^2} (\bar{k}_{x}-k_{x}) \tr \left[\bm{\alpha}(\omega)\cdot \imaginary\, \bm{g}'(\omega,\bm{k}_{\perp};a,a)\cdot\bm{\alpha}(\omega)\cdot \imaginary\, \bm{g}'(\omega,\bar{\bm{k}}_{\perp};a,a)\right]\coth\frac{\beta\gamma(\omega+\bar{k}_{x}v)}{2},
\end{equation}
where $ \gamma=1/\sqrt{1-v^{2}} $ is the relativistic dilation factor. The atom's polarizability, $ \bm{\alpha} $, is defined in its own rest frame, $ \mathcal{P} $, while the inverse temperature, $ \beta $, is defined in the rest frame of the radiation, $ \mathcal{R} $. The $ \bm{g}' $ in Eq.~\eqref{starting} is not the usual reduced Green's dyadic, $ \bm{g} $, defined in frame $ \mathcal{R} $, but, rather, its Lorentz transform to frame $ \mathcal{P} $. Each component of $ \bm{g'} $ can be expressed in terms of a combination of different components of $ \bm{g} $, which we detail in Appendix \ref{gprime}. The explicit form of $ \bm{g} $ for a general planar background is recorded in Appendix \ref{g}. Finally, $ \imaginary \bm{g}$ refers to the anti-Hermitian part of the Green's dyadic, of which the components are
\begin{equation}
(\imaginary \bm{g})_{ij}(\omega,\bm{k}_{\perp}; z,\tilde{z})=\frac{g_{ij}(\omega,\bm{k}_{\perp}; z,\tilde{z})-g_{ji}^{*}(\omega,\bm{k}_{\perp}; \tilde{z},z)}{2i} =\frac{g_{ij}(\omega,\bm{k}_{\perp}; z,\tilde{z})-g_{ji}(-\omega,-\bm{k}_{\perp}; \tilde{z},z)}{2i}.
\end{equation}

The quantum friction in Eq.~\eqref{starting} is in fact the $ x $ component of the Lorentz force on a moving dipole quantized using the fluctuation-dissipation theorem (FDT). Because the atom is intrinsically nondissipative, the frictional force is second order in $ \bm{\alpha} $ like that discussed in Ref.~\cite{Xin:eqf1}. There are two contributions to the force: the $ \bar{k}_{x} $ term comes from the field fluctuations directly while the $ k_{x} $ term comes from the induced dipole fluctuations. Although entering the friction formula with different signs, the two contributions do not cancel each other due to the Doppler shifting of the frequency in the $ \coth $ factor. For an ordinary point particle, the polarizability of which is reciprocal, a relative velocity between the particle and the surrounding blackbody background is necessary for any transverse force to arise.\footnote{For a nonreciprocal point particle which is not in thermal equilibrium with the environment,  self propulsion can be induced even if the particle is initially at rest. We will explore the fluctuation-induced effects for such a nonreciprocal particle in Ref.~\cite{Kim:torque}. In addition, self propulsion is also possible for extended objects made up of reciprocal but nonuniform materials. See, for example, Ref.~\cite{Reid:photon} by Reid et al.} Even at zero temperature, the quantum friction does not vanish in general. But we have learned from our previous investigations that if the background is just free space, the resultant QVF does vanish at zero temperature \cite{Xin:eqf1, Xin:eqf2}. 

The matrix structure under the trace in the integrand is in general complicated. Even for an isotropic atom, there will be contributions to the quantum friction that mix the different diagonal polarization states of the atom and pick up the off-diagonal components of $ \bm{g}' $.\footnote{The effective polarizability, $ \hat{\bm{\alpha}}= \bm{\alpha}\cdot\bm{\Gamma}\cdot{\bm{\alpha}}$, as defined in Ref.~\cite{Xin:eqf1}, can acquire off-diagonal components through the off-diagonal components of the Green's dyadic, even when the intrinsic polarizability of the atom, $ \bm{\alpha} $, is diagonal.} Moreover, each component of $ \bm{g}' $ is still to be re-expressed as a combination of the different components of $ \bm{g} $ in the actual calculation. However, the special background indicated in Fig.~\ref{setup} has several features which greatly help to simplify the calculation. In the presence of the PC plate, $ \bm{g} $ is found using the general expressions in Appendix \ref{g} to be
\begin{align}\label{gPC}
\bm{g}^{\rm{PC}}(\omega,\bm{k}_{\perp};a,a)=
&\mqty(\frac{\omega^{2}-k_{x}^{2}}{2\kappa}(1-e^{-2\kappa a})\qquad
&
\;-\frac{k_{x}k_{y}}{2\kappa}(1-e^{-2\kappa a})
&
-\frac{i}{2}k_{x}e^{-2\kappa a}\qquad
\\\\
-\frac{k_{x}k_{y}}{2\kappa}(1-e^{-2\kappa a})
&
\;\frac{\omega^{2}-k_{y}^{2}}{2\kappa}(1-e^{-2\kappa a})\qquad
&
-\frac{i}{2}k_{y}e^{-2\kappa a}\qquad
\\\\
+\frac{i}{2}k_{x}e^{-2\kappa a}\qquad
&
\quad +\frac{i}{2}k_{y}e^{-2\kappa a}\qquad
&
\frac{k^{2}}{2\kappa}(1+e^{-2\kappa a}))
\end{align}
with $ \kappa=\sqrt{k^{2}-\omega^2} $.
Similar to the vacuum background, the permittivity of a PC plate, $ \varepsilon\to\infty  $, is invariant under a Lorentz transformation in the $ x $ direction. As a result, $ \bm{g}' $ is found to be identical to $ \bm{g} $ when applying the transformations listed in Appendix \ref{gprime}:
\begin{equation}\label{surprise}
\bm{g'}^{\rm{PC}}(\omega,\bm{k}_{\perp};a,a)=\bm{g}^{\rm{PC}}(\omega,\bm{k}_{\perp};a,a).
\end{equation}
That is, for this special case, we can replace $ \bm{g}' $ in Eq.~\eqref{starting} with $ \bm{g} $ in Eq.~\eqref{gPC}. In addition, terms containing the product of $ g_{xy} $ and $ g_{yx} $ or $ g_{yz} $ and $ g_{zy} $ do not contribute to the friction due to their oddness in $ k_{y} $. If the polarizability tensor of the atom is diagonal, the only contribution to the friction involving the off-diagonal components of the Green's dyadic, which mixes different components of the polarizability tensor is
\begin{equation}\label{FXZ}
F^{\rm{XZ}}=2\int\frac{d\omega}{2\pi}\frac{d^{2}\bm{k}_{\perp}}{(2\pi)^{2}}\frac{d^{2}\bar{\bm{k}}_{\perp}}{(2\pi)^{2}}\, (\bar{k}_{x}-k_{x}) \alpha_{xx}(\omega)(\imaginary \bm{g})_{xz}^{\rm{PC}}(\omega,\bm{k}_{\perp};a,a)\alpha_{zz}(\omega)(\imaginary\bm{g})_{zx}^{\rm{PC}}(\omega,\bar{\bm{k}}_{\perp};a,a)\coth\left[\frac{\beta}{2}\gamma(\omega+\bar{k}_{x}v)\right].
\end{equation}
Here and in the rest of the paper, we use a superscript on $ F $ to specify contributions from different  polarization states of the atom. In fact, $ F^{\rm{XZ}} $ turns out to be the most interesting contribution to the frictional force, because it actually corresponds to a push instead of a drag.

Before presenting the results we obtain for QFPC, let us stress that no QFPC arises at zero temperature. In Appendix C, we prove that the zero temperature QF is absent not only for the vacuum case and the PC case, but also for the broader class of diaphanous materials when a nontrivial permeability ($ \mu\neq 1 $) is taken into account.

\section{EXACT RESULTS AND VARIOUS LIMITS}
We see from the starting formula Eq.~\eqref{starting} that each contribution to the frictional force is proportional to the product of two nonvanishing components of the polarizability tensor. For simplicity, let us assume the polarizability tensor to be diagonal. We have ruled out most of the contributions that mix the components of the polarizability tensor based on the symmetry of the integrand. Then there are only four nonvanishing contributions left. They are proportional to $ \alpha_{xx}^2 $, $ \alpha_{yy}^2 $, $ \alpha_{zz}^2 $, $ \alpha_{xx}\alpha_{zz} $ and will be denoted as $ F^{\rm{XX}} $, $F^{\rm{YY}} $, $F^{\rm{ZZ}} $ and $F^{\rm{XZ}} $, respectively. 

Crucial to the calculation is finding the anti-Hermitian part of the relevant components of $ \bm{g}^{\rm{PC}} $. It can be seen from Eq.~\eqref{gPC} that $ \imaginary \,\bm{g}^{\rm{PC}}=\bm{0} $ unless the propagation wave number $ \kappa $ develops an imaginary part. Since the integrand in Eq.~\eqref{starting} involves the product of two Green's dyadics evaluated at $ (\omega, \bm{k}_{\perp}) $ and $ (\omega,\bar{\bm{k}}_{\perp}) $, respectively, the integration is restricted to regions where the propagation wave numbers associated with both Green's dyadics become imaginary,
\begin{equation}\label{regions}
\kappa\to -i\sgn(\omega)\sqrt{\omega^{2}-k^{2}}, \quad k^2<\omega^2, \qquad \bar{\kappa}\to -i\sgn(\omega)\sqrt{\omega^{2}-\bar{k}^{2}}, \quad \bar{k}^2<\omega^2.
\end{equation}
The branches need to be chosen so that the Green's dyadic is retarded. The anti-Hermitian parts of the relevant components of the Green's dyadic therefore read
\begin{subequations}\label{Img}
\begin{equation}
(\imaginary\bm{g})_{xx}(\omega,\bm{k}_{\perp}; a, a)=\Im\, g_{xx}(\omega,\bm{k}_{\perp}; a, a)=\sgn(\omega) \frac{\omega^2-k_{x}^2}{2\sqrt{\omega^2-k^2}}\left[1-\cos(2\sqrt{\omega^2-k^2}a)\right],
\end{equation}
\begin{equation}
(\imaginary\bm{g})_{yy}(\omega,\bm{k}_{\perp}; a, a)=\Im\, g_{yy}(\omega,\bm{k}_{\perp}; a, a)=\sgn(\omega) \frac{\omega^2-k_{y}^2}{2\sqrt{\omega^2-k^2}}\left[1-\cos(2\sqrt{\omega^2-k^2}a)\right],
\end{equation}
\begin{equation}
(\imaginary\bm{g})_{zz}(\omega,\bm{k}_{\perp}; a, a)=\Im\, g_{zz}(\omega,\bm{k}_{\perp}; a, a)=\sgn(\omega) \frac{k^2}{2\sqrt{\omega^2-k^2}}\left[1+\cos(2\sqrt{\omega^2-k^2}a)\right],
\end{equation}
\begin{equation}
(\imaginary\bm{g})_{xz}(\omega,\bm{k}_{\perp}; a, a)=-(\imaginary\bm{g})_{zx}(\omega,\bm{k}_{\perp}; a, a)=-i\sgn(\omega)\frac{k_{x}}{2}\sin(2\sqrt{\omega^2-k^2}a).
\end{equation}
\end{subequations}
The off-diagonal components of $ \imaginary\bm{g} $ are different from the diagonal components in several respects. First of all, they are purely imaginary.  Second, they are odd in $ k_{x} $. As a result, in Eq.~\eqref{starting}, $ F^{\rm{XZ}} $ contributes to the total friction through the $ -k_{x} $ term, while $ F^{\rm{XX}} $, $ F^{\rm{YY}} $ and $ F^{\rm{ZZ}} $ all contribute through the $ \bar{k}_{x} $ term. It is precisely the apparent minus sign in the $ -k_{x} $ term that renders $ F^{\rm{XZ}} $  positive, corresponding to a push instead of a drag.\footnote{Physically, the $ -k_{x} $ term comes from the induced dipole fluctuations, while the $ \bar{k}_{x} $ term comes from the direct field fluctuations. In the QVF case, these two different contributions also occur and they are called $ F_{\rm{II}}' $ and $ F_{\rm{I}}' $, respectively, in Ref.~\cite{Xin:eqf1}. There, $ F_{\rm{II}}' $ vanishes, which reflects the fact that the dipole radiation emitted by the atom is isotropic. Here, the fact that $ F^{\rm{XZ}} $ exists through the $ -k_{x} $ term indicates that the dipole radiation in the $ x $ direction is no longer isotropic when the PC plate is present. Furthermore, the positive sign of this contribution reflects that the corresponding dipole radiation emitted backward must exceed that emitted forward.} Third, they do not contain terms independent of the atom-plate separation, $ a $, as those in the diagonal components. These terms reflect the vacuum contributions. So, the off-diagonal components do not contribute to the QVF discussed in Ref.~\cite{Xin:eqf1}.

Without any further assumptions, we insert Eq.~\eqref{Img} into Eq.~\eqref{starting} and integrate  $ k_{x} $, $ k_{y} $ and $ \bar{k}_{y} $ analytically. With a further change of variable, $ \bar{k}_{x}=\omega u $, we find the contribution to the QFPC from the $ \rm{PQ} $ polarization states can be written as
\begin{equation}\label{summary}
F^{\rm{PQ}}=\frac{1}{32\pi^{3}}\int_{0}^{\infty} \,d\omega\, \alpha_{pp}(\omega)\alpha_{qq}(\omega)\, \omega^{7} \mathcal{F}^{\rm{PQ}}(x,v,z),
\end{equation}
and for each contribution, $ \mathcal{F}^{\rm{PQ}} $ reads
\begin{subequations}
\label{main}
\begin{align}
\label{mainX}
\mathcal{F}^{\rm{XX}}(x,v,z)=&\left\lbrace \frac{4}{3}-\frac{2}{x^{3}}\left[x\cos x+(x^{2}-1)\sin x\right]\right\rbrace\int_{-1}^{1} du\, u\left(1-u^{2}\right)\left[1-J_{0}\left(x\sqrt{1-u^2}\right)\right]\frac{1}{e^{x\gamma(1+uv)z}-1},
\end{align}
\begin{align}
\label{mainY}
\mathcal{F}^{\rm{YY}}(x,v,z)=&\left\lbrace \frac{4}{3}-\frac{2}{x^{3}}\left[x\cos x+(x^{2}-1)\sin x\right]\right\rbrace\nonumber\\\cross &\int_{-1}^{1} du \, u\left[\frac{1}{2}(1+u^2)-J_{0}\left(x\sqrt{1-u^2}\right)+\frac{\sqrt{1-u^2}}{x}J_{1}\left(x\sqrt{1-u^2}\right)\right]\frac{1}{e^{x\gamma(1+uv)z}-1},
\end{align}
\begin{align}
\label{mainZ}
\mathcal{F}^{\rm{ZZ}}(x,v,z)=&\left\lbrace \frac{4}{3}-\frac{4}{x^{3}}\left[x\cos x-\sin x\right]\right\rbrace\nonumber\\
\cross &\int_{-1}^{1} du\, u\left[\frac{1}{2}(1+u^2)+u^2 J_{0}\left(x\sqrt{1-u^2}\right)+\frac{\sqrt{1-u^2}}{x}J_{1}\left(x\sqrt{1-u^2}\right)\right]\frac{1}{e^{x\gamma(1+uv)z}-1},
\end{align}
\begin{align}
\label{mainXZ}
\mathcal{F}^{\rm{XZ}}=&-2\left\lbrace \frac{2}{x^{4}}\left[-3x\cos x-(x^{2}-3)\sin x\right]\right\rbrace\int_{-1}^{1} du \,u\sqrt{1-u^2}J_{1}\left(x\sqrt{1-u^2}\right)\frac{1}{e^{x\gamma(1+uv)z}-1}.
\end{align}
\end{subequations}

Here, we have introduced a dimensionless frequency scaled by the distance $ a $, 
\begin{equation}
 x=2\omega a, 
\end{equation}
as well as a dimensionless inverse temperature also scaled by $ a $,
\begin{equation}
z=\frac{\beta}{2a}=\frac{1}{2aT}.
\end{equation}

So far, the expressions we have for QFPC in Eq.~\eqref{main} are exact and involve the dynamical polarizability of the atom. For frequencies smaller than the lowest excitation energy of the atom, the dynamical polarizability, $ \bm{\alpha}(\omega) $, can be replaced by the static polarizability \cite{Miller1978}, $ \bm{\alpha}(0) $. Due to the common exponential factors in Eq.~\eqref{mainX}--\eqref{mainXZ}, the high frequency modes with $ \beta\omega=x z\gg 1 $ will be cut off and do not significantly contribute to the $\omega$ integral. Therefore, so long as the temperature is not high enough to excite the atom to its higher energy states, we can work in the static limit, where we substitute the polarizability with its static value. This allows us to take the product of the polarizabilities out of the $ \omega $ integral in Eq.~\eqref{summary}:
\begin{equation}\label{scale}
F^{\rm{PQ}}=\frac{\alpha_{pp}(0)\alpha_{qq}(0)}{32\pi^{3}(2a)^{8}}f^{\rm{PQ}}(v,z), \quad f^{\rm{PQ}}(v,z)=\int_{0}^{\infty} dx\, x^{7} \mathcal{F}^{\rm{PQ}}(x,v,z),
\end{equation}
where the dimensionless functions $ f^{\rm{PQ}} $ now characterize contributions to QFPC from different polarization states.

Note the magnitude of $ z $ determines the dominating modes of the $ x $ integral in Eq.~\eqref{scale}. For $ z\ll 1 $, it is dominated by the large $ x $ modes, where the complicated $ x $ dependences in the integrands become subdominant and drop out, except for the common factor of $ \frac{x^7}{e^{x\gamma(1+uv)z}-1} $. As a result, the diagonal contributions $ F^{\rm{XX}} $, $ F^{\rm{YY}} $ and $ F^{\rm{ZZ}} $ become distance independent and proportional to $ T^{8} $. Indeed, for $ z\ll 1 $, the diagonal contributions of QFPC precisely reduce to the corresponding contributions of QVF in Ref.~\cite{Xin:eqf1}. On the other hand, $ F^{\rm{XZ}} $, which is proportional to $ T^{4}/a^{4} $, becomes completely negligible in comparison to the diagonal contributions. To sum up, the contributions to QFPC in the small $ z $ limit read
\begin{equation}
\label{vaclim}
F^{\rm{PQ}}_{z\ll 1}=\frac{\alpha_{pp}^2(0)}{32\pi^{3}(2a)^{8}}f_{z\ll 1}^{\rm{PQ}}(v,z),\quad f_{z\ll 1}^{\rm{PQ}}(v,z)=\addtolength{\arraycolsep}{-3pt}
\left\{%
 \begin{array}{lcrcl}
 -\frac{4\Gamma(8)\zeta(8)}{3z^{8}}\frac{32}{105}\gamma^4 v(7+3v^2),& \qquad \rm{PQ}=\rm{XX}\\\\
-\frac{4\Gamma(8)\zeta(8)}{3z^{8}}\frac{32}{105}\gamma^6 v(14+37v^2+9v^4),& \qquad \rm{PQ}=\rm{YY}, \rm{ZZ}\\\\
\frac{16\Gamma(4)\zeta(4)}{z^4}\frac{v}{\gamma^4},& \qquad \rm{PQ}=\rm{XZ}.
 \end{array}
 \right.
\end{equation}
As is shown in Eq.~\eqref{vaclim}, unlike the diagonal contributions which monotonically increase with velocity, we find $ F^{\rm{XZ}}$ vanishes when the velocity approaches the speed of light. Since the small $ z $ limit of $ F^{\rm{XZ}} $ has not been worked out in Ref.~\cite{Xin:eqf1}, we show how to obtain it analytically in Appendix \ref{limit}. 

It is not so surprising that the small $ z $ limit of QFPC coincides with QVF. Small $ z $ values correspond to large distances or high temperatures. When the atom is far away from the PC plate, it is obvious that QFPC should reduce to QVF. In the case of high temperatures (but not so high to ionize the atom), the atom interacts with photons of very high frequency. It therefore mainly probes the very short distances around it and, effectively, does not feel the PC plate. That is, in the high temperature limit, the distribution of energy eigenvalues of photons interacting with the atom is insensitive to the presence of the plate. 

Since quantum vacuum friction has been explored for a nondissipative atom in Ref.~\citep{Xin:eqf1}, the new physics really lies in the large $ z $ limit, the short-distance or low-temperature behavior of QFPC. For $ z\gg 1 $, the small $ x $ modes dominate the integrals. We can therefore expand the integrands in powers of $ x $ before carrying out the integrals. Quite interestingly, the integrands for various polarization states exhibit different leading power behavior in $ x $, which determines the distance and temperature dependences of their contributions to QFPC. After expansion in $ x $, both the $ x $ and $ u $ integrals can be done exactly if we keep only the leading in $z$ terms. (Appendix \ref{limit} contains an approach to derive the asymptotic expression for $ F^{\rm{XZ}} $ in the large $ z $ limit as well.) For $ z\gg 1 $, the resultant QFPC is found to be
\begin{equation}
\label{key}
F^{\rm{PQ}}_{z\gg 1}=\frac{\alpha_{pp}(0)\alpha_{qq}(0)}{32\pi^{3}(2a)^{8}}f_{z \gg 1}^{\rm{PQ}}(v,z),
\quad 
f_{z\gg 1}^{\rm{PQ}}(v,z)=\addtolength{\arraycolsep}{-3pt}
\left\{%
 \begin{array}{lcrcl}
 -\frac{\Gamma(12)\zeta(12)}{15z^{12}}\frac{64}{3465}\gamma^6 v(99+110v^2+15v^4),& \quad \rm{PQ}=\rm{XX}\\\\
 -\frac{\Gamma(12)\zeta(12)}{15z^{12}}\frac{32}{3465}\gamma^8 v(297+1034v^2+625v^4+60v^6),& \quad \rm{PQ}=\rm{YY}\\\\
-\frac{8\Gamma(8)\zeta(8)}{3z^{8}}\frac{64}{105}\gamma^6 v(14+37v^2+9v^4),& \quad \rm{PQ}= \rm{ZZ}\\\\
2\frac{\Gamma(10)\zeta(10)}{15z^{10}}\frac{8}{63}\gamma^6 v(21+30v^2+5v^4),& \quad \rm{PQ}= \rm{XZ}.
 \end{array}
 \right.
\end{equation}
Since the results shown in Eq.~\eqref{key} are for the large $ z $ limit, it is apparent that $ F^{\rm{ZZ}} $ dominates over the contributions from the other polarizations. In this limit, $ F^{\rm{ZZ}} $ is independent of distance $ a $ and proportional to $ T^8 $, just as is the case for QVF. In fact, we find $ F^{\rm{ZZ}} $ is precisely four times the corresponding QVF contribution shown in Eq.~\eqref{vaclim}. The next leading contribution, $ F^{\rm{XZ}} $, is proportional to $ a^2 T^{10} $ with an overall positive sign, suggesting that this particular contribution corresponds to a push instead of a drag. The smallest contributions, $ F^{\rm{XX}} $ and $ F^{\rm{YY}}$, are both proportional to $ a^4 T^{12} $. On closer examination of Eq.~\eqref{key}, we also observe that $ f^{\rm{YY}}$ is always greater than $f^{\rm{XX}} $, for arbitrary velocities. 

Interestingly, these behaviors of QFPC may be easily understood from the image particle picture criticized in the Introduction. In fact, there is nothing wrong with replacing the PC plate by an image particle moving synchronously with the actual particle. We only need to keep in mind that both particles would interact with the surrounding photon bath, so that a frictional force does indeed arise. Following this line of reasoning, the image particle would double the normal component of the fluctuation-induced field, $ E_{z} $, but eliminate the tangential  components, $ E_{x} $ and $ E_{y} $, at the surface of the PC plate. Since these fluctuation-induced frictional forces are proportional to the product of the relevant fields, $ F^{\rm{ZZ}} $ is therefore quadrupled while the other contributions are all suppressed when the distance between the particle and the PC plate approaches zero. 

We have been advised by Matthias Kr\"{u}ger that the physics here is analogous to a classical situation in hydrodynamics. For example, the authors of Ref.~\cite{Kruger:colloid} studied colloidal particles driven through a suspension of mutually noninteracting Brownian particles and the corresponding frictional force induced by the nonequilibrium fluid structure. (The flow field comoving with the colloidal particles is not in equilibrium with the Brownian particles.) They found that the frictional force on a single colloidal particle traveling along a wall (analogous to the PC plate in our case) is precisely the same as that on two colloidal particles driven side by side. The authors also found an enhancement of the friction due to the wall/image colloidal particle in comparison to the friction on an isolated colloidal particle. From the density plot of the solute Brownian particles, they interpret this increase in friction as the result of more Brownian particles aggregating in front of the colloidal particles when the wall/image particle is present. An analogous interpretation applies to what we see here in this paper. That is, the electromagnetic energy density is stronger near the PC plate.
 
So far, both the small $ z $ results in Eq.~\eqref{vaclim} and the large $ z $ results in Eq.~\eqref{key} are exact in velocity. Another question is whether we can obtain the nonrelativistic (NR) limit analytically without assuming anything about $ z $. This is possible as long as the dynamical polarizability is still replaced by its static value. We illustrate the procedure of obtaining the NR limit for $ F^{\rm{XZ}} $, valid for all $ z $ values in Appendix \ref{NR}. It turns out that all contributions to QFPC start with a term linear in $ v $ in the NR limit. 

\section{NUMERICAL RESULTS}
As one of the contributions, $ F^{\rm{XZ}} $, is positive (a push), while the others are all negative (a drag), a natural question arises: could the overall ``frictional" force on an atom ever flip sign and therefore become a push? Of course, from Eq.~\eqref{vaclim} and Eq.~\eqref{key}, we can already conclude that the overall QFPC is negative definite in both the small $z$ (vacuum/high-temperature) limit  and large $ z $ (short-distance/low-temperature) limit. But, there is no convincing argument just from the analytic results suggesting that QFPC cannot switch sign in the intermediate $ z $ regime. Therefore, we resort to numerical methods to ascertain the sign of QFPC.

We will here mainly consider atoms in their ground states, the polarizability of which is normally quite isotropic\footnote{Closed-shell atoms are almost exactly isotropic \cite{polarizability-table}. Even for open-shell atoms, the anisotropy is typically small compared to the isotropic part of the polarizability. Among the elements in a certain period, the anisotropy is largest when the first $ p $ electron is added \cite{Miller1978}. } and can be well approximated by its static value, $ \bm{\alpha}(\omega)=\alpha(0)\bm{1} $. For such isotropic atoms, the sign of the QFPC is determined by the sum of the dimensionless functions introduced in Eq.~\eqref{scale}:
\begin{equation}\label{iso}
F^{\rm{ISO}}=\frac{\alpha^{2}(0)}{32\pi^{3}(2a)^{8}}f^{\rm{ISO}}(v,z),\quad f^{\rm{ISO}}(v,z)=f^{\rm{XX}}(v,z)+f^{\rm{YY}}(v,z)+f^{\rm{ZZ}}(v,z)+f^{\rm{XZ}}(v,z).
\end{equation}

\begin{figure}[h!]
 \includegraphics[width=0.70\linewidth]{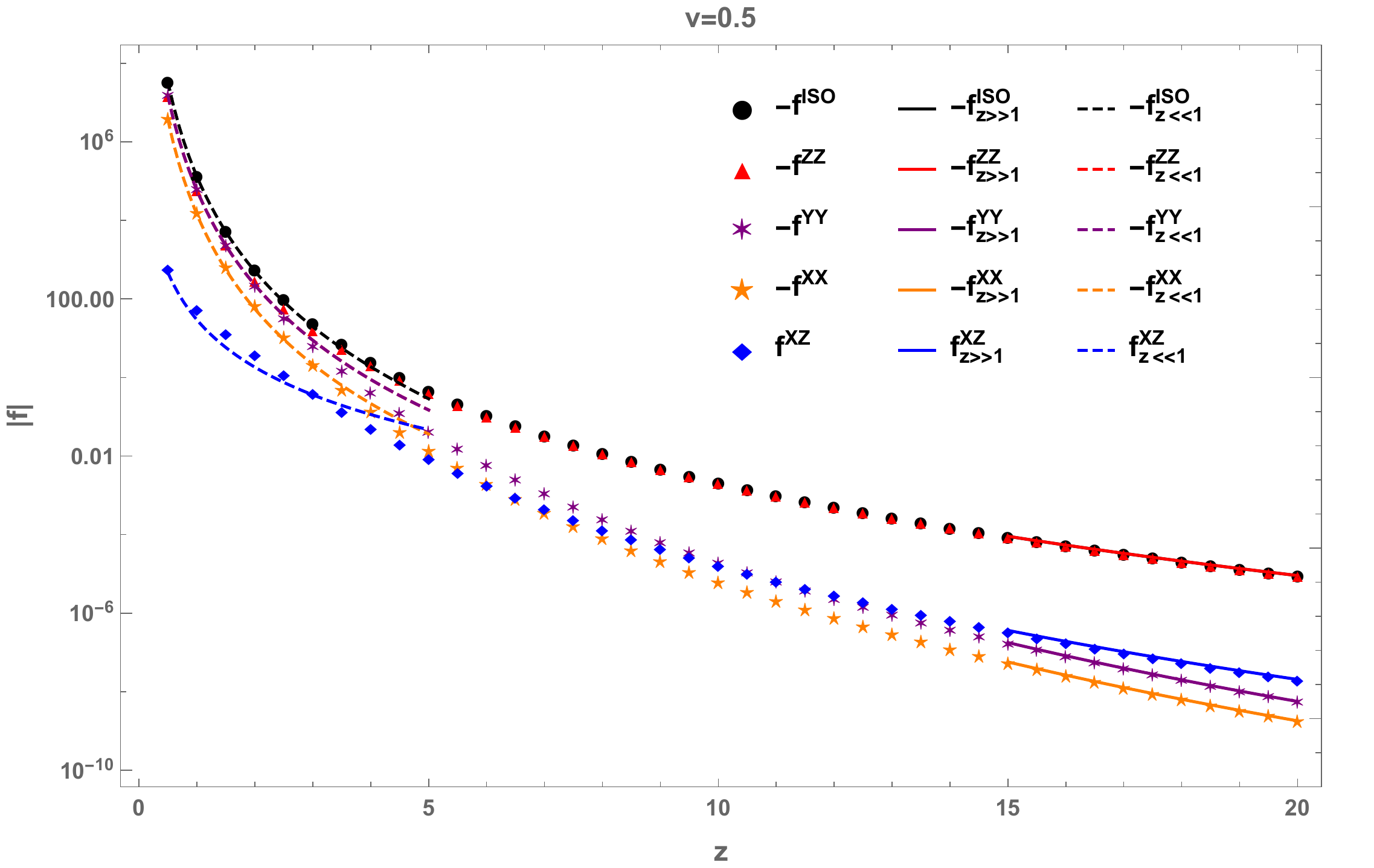}
 \caption{The absolute values of the dimensionless functions $ f^{\rm{PQ}} $ in Eq.~\eqref{iso} are shown as functions of $ z $ for fixed velocity ($ v=0.5 $). The numerical results are computed directly using Eq.~\eqref{main} and Eq.~\eqref{scale}. Their small $ z $ and large $ z $ approximations are obtained from Eq.~\eqref{vaclim} and Eq.~\eqref{key}, respectively. Since the small $ z $ approximation for $ f^{\rm{ZZ}} $ and $ f^{\rm{YY}} $ is identical, the dashed purple line overlaps with the dashed red line. As is seen, the small $ z $ approximation of $ F^{\rm{XZ}} $ cannot give a good description of the numerical data beyond $ z=1 $. A further detailed plot is provided in Appendix \ref{limit}, where the agreement between the analytic approximation and the numerical data for $ F^{\rm{XZ}} $ is more clearly demonstrated for smaller $ z $ values. } \label{Fig2}
 \end{figure}

We show the absolute value of these dimensionless functions across their transition region in Fig.~\ref{Fig2}. Starting from small $ z $ values, the total frictional force on the isotropic particle is dominated almost evenly between the $ \rm{ZZ} $ and $ \rm{YY} $ contributions. But as $ z $ grows larger, the weight of the $ \rm{YY} $ contribution decays so that the $ \rm{ZZ} $ contribution solely dominates the entire frictional force. As for the unique positive contribution from the $ \rm{XZ} $ polarization, it is completely negligible when $ z $ is small but it eventually surpasses the contributions from the $ \rm{XX} $ and $ \rm{YY} $ polarizations for large $ z $. Nonetheless, it never dominates the $ \rm{ZZ} $ polarization. The asymptotic (in $ z $) expressions in Eq.~\eqref{vaclim} and Eq.~\eqref{key} are consistent with these behaviors and the agreement with the numerical data in their supposedly valid regimes are also clearly illustrated in the figure. So, we can conclude that the total QFPC on an isotropic atom \emph{is always a drag}, since it \emph{cannot} change sign even in the intermediate $ z $ regime. 
 % \begin{figure*}[h!]
%\subfloat[]{\label{fig1a}%
%\includegraphics[width=0.50\linewidth]{Fig1a(10)}%
%}
%\subfloat[]{\label{fig1b}%
%\includegraphics[width=0.48\linewidth]{Fig1b(2)}%
%}
%\caption{}
%\label{Fig1}
%\end{figure*}
%In order to see the transition behavior more clearly, we plot the ratio $ r^{\rm{PQ}}=f^{\rm{PQ}}/f^{\rm{ISO}}$ in Fig.~\ref{Fig3}. Initially, for smaller $ z $ values, $ f^{\rm{ISO}} $ is dominated by $ f^{\rm{ZZ}} $ and $ f^{\rm{YY}} $ almost evenly. This is just what is expected to be the case in the vacuum limit. And gradually, as $ z $ grows larger, the weight of $ f^{\rm{YY}} $ decays while $ f^{\rm{ZZ}} $ takes over completely.
%\begin{figure}[h!]
% \includegraphics[width=0.7\linewidth]{Fig3(2)}
% \caption{The weight of each individual contribution in the total force, $ r^{\rm{PQ}} $, is shown as a funciton of $ z $ for a fixed velocity $ v=0.5 $. Note the values for $ r^{\rm{XZ}} $ are all negative with a tiny magnitude.} \label{Fig3}
% \end{figure} 

\begin{figure}
 \includegraphics[width=0.55\linewidth]{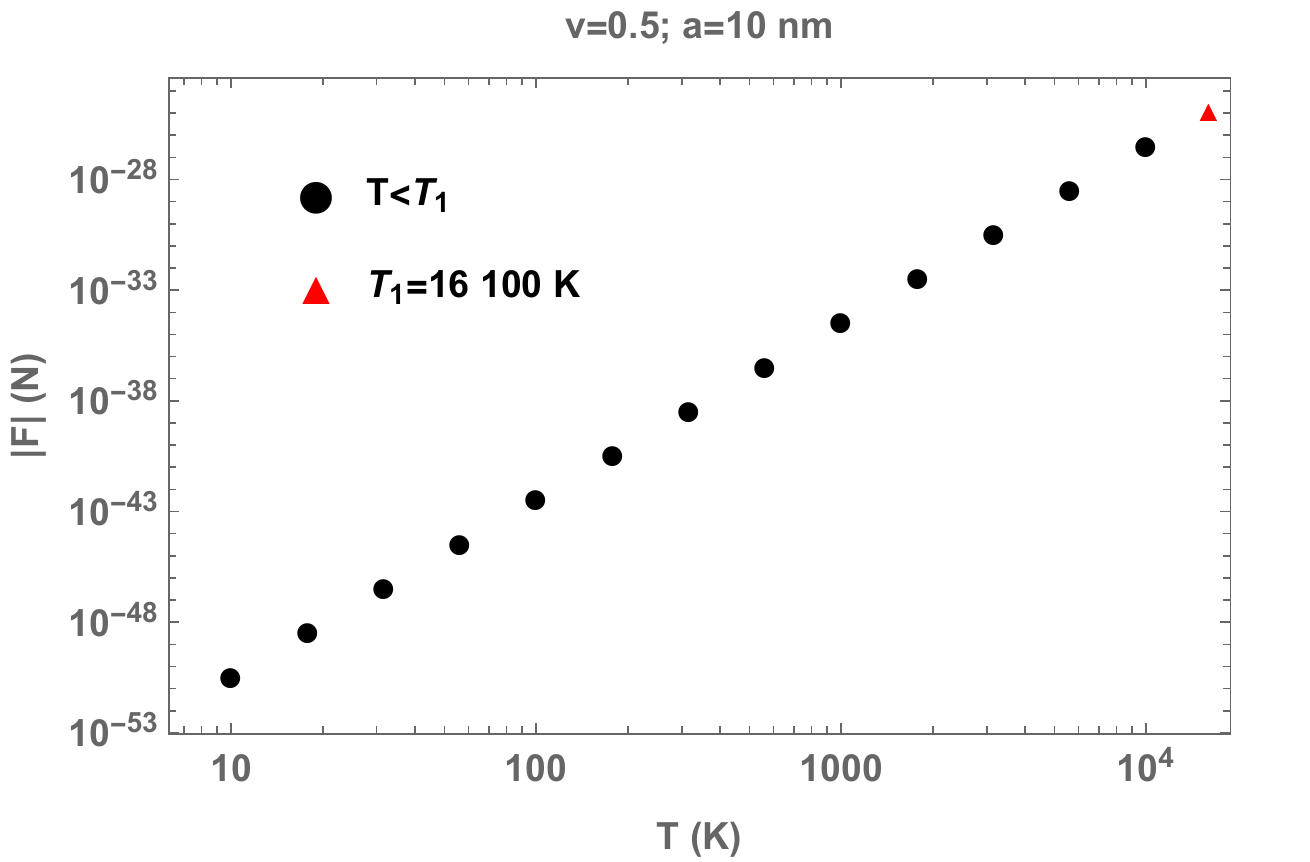}
 \caption{The magnitude of the total frictional force on a Cs atom moving at $ v=0.5 $ and at a distance $ a=10 \,\rm{nm} $ from the PC plate is plotted as a function of temperature. The  friction at the first excited temperature is indicated by the red triangle, with a magnitude of $ 1.30\cross 10^{-25}\, \rm{N} $. } \label{FT}
 \end{figure}

Another interesting aspect of the force is, of course, its magnitude. Fluctuation induced forces are typically small. But, is the QFPC possibly accessible to experiment?  
Here, we estimate QFPC on a cesium (Cs) atom, which has the largest static polarizability,\footnote{Within a period, the alkali metal atoms generally have the biggest polarizabilities. They are also supposed to have very tiny anisotropy because their valence electrons are in $ s $ states \cite{Miller1978}. Cs has the largest polarizability among the alkali metal atoms.} according to Ref.~\cite{polarizability-table}, $ \alpha_{\rm{Cs}}(0)= 59.3 \, \mathrm{\mathring{A}}^{3}$. Because the expression \eqref{scale} we use for numerical calculation is obtained in the static limit, the corresponding numerical results are only expected to be appropriate when the atom is in its ground state, that is, up to the temperature that corresponds to the \emph{first excitation energy} of the Cs atom, $ T_{1}=16\,100 \, \rm{K} $,\footnote{This temperature and the ionization temperature used later are obtained from the first excitation energy of Cs listed in Ref.~\cite{Weber:csenergy}.} beyond which a model for its dynamical polarizability is needed. In Fig.~\ref{FT}, we show the magnitude of the total frictional force on a Cs atom  up to $ T_{1} $, fixing velocity and distance. The friction clearly exhibits a power-law dependence on temperature. This is no surprise because we already know that 
the frictional force should behave as $ T^8 $ in both the large $ z $ (low $ T $) and small $ z $ (high $ T $) regimes.

\begin{figure}[h]
 \includegraphics[width=0.65\linewidth]{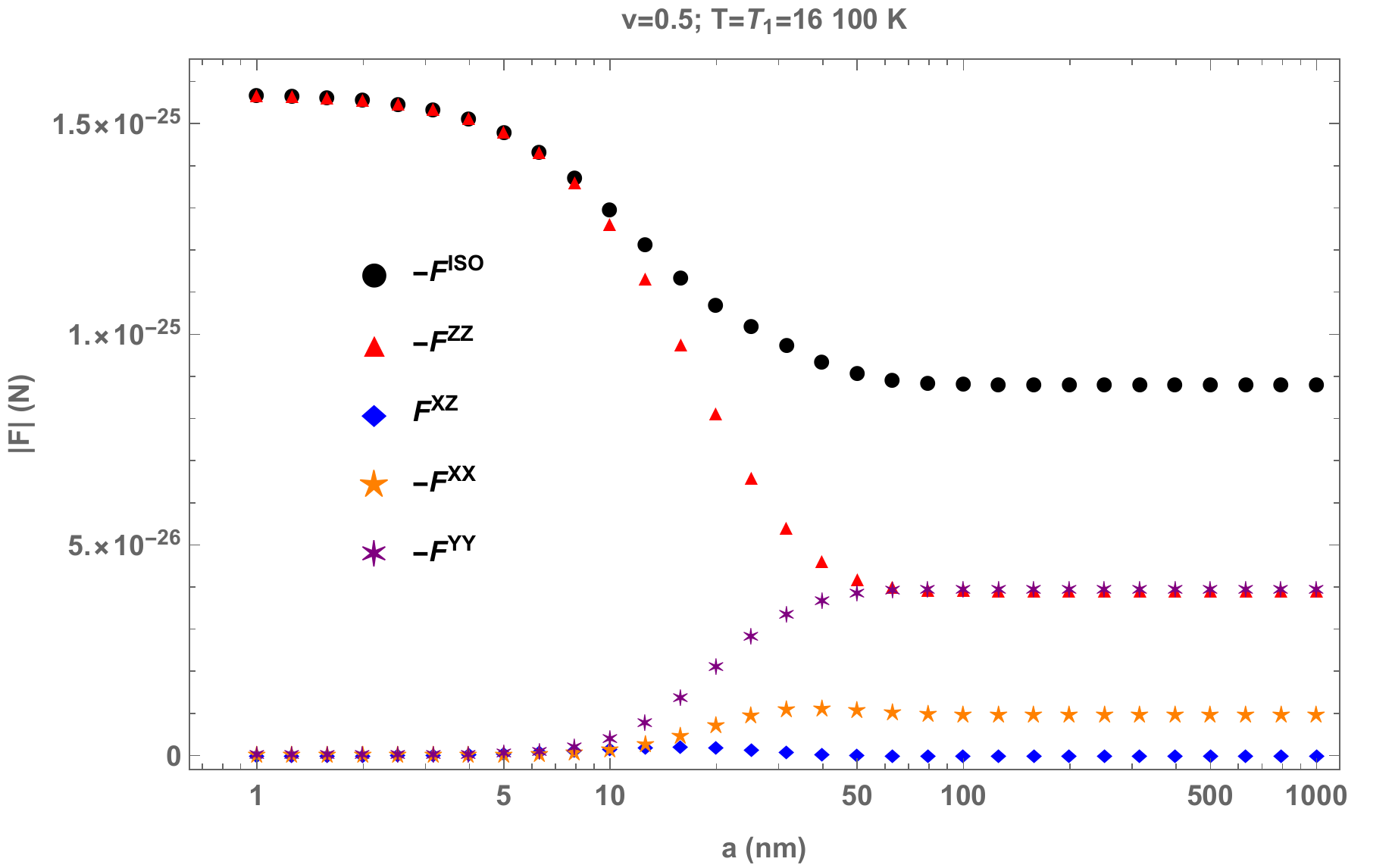}
 \caption{The magnitude of the total frictional force, along with its contributions from different polarizations, on a Cs atom moving at $ v=0.5 $ and at its first excited temperature $ T=16\,100 \,\rm{K} $ is plotted as a function of distance. The largest magnitude of the total friction shown by the black dots is for $ a=1\,\rm{nm} $, being $ 1.57\cross 10^{-25}\, \rm{N} $.  } \label{Fa}
 \end{figure}

Of course, QFPC also depends on the distance between the atom and the plate, distinguishing it from QVF. Considering the size of the Cs atom\footnote{Cesium also has the largest covalent  radius ($ 244\,\rm{pm} $) among the nonradioactive atoms according to Ref.~\cite{Covalent:radii}.}, we should keep the distance greater than $ 1\,\rm{nm} $ to avoid additional surface effects. We therefore show the magnitude of QFPC for a Cs atom as a function of distance in Fig.~\ref{Fa}, from  $ 1 \,\rm{nm} $ to $1 \,\rm{\mu m}$, fixing the velocity at $ v=0.5 $ and temperature at $ T=T_{1} $. It is seen that the total friction is only doubled when the distance is reduced from $ 1\,\rm{\mu m} $ to $ 1\,\rm{nm} $. This can be well understood from the asymptotic behavior of the dominant contributions: the $ \rm{ZZ} $ contribution quadruples, yet the $ \rm{YY} $ contribution vanishes at small distances, which is also clearly illustrated in the figure.
 
\begin{figure*}[h]
\subfloat[]{\label{FvNR}%
\includegraphics[width=0.49\linewidth]{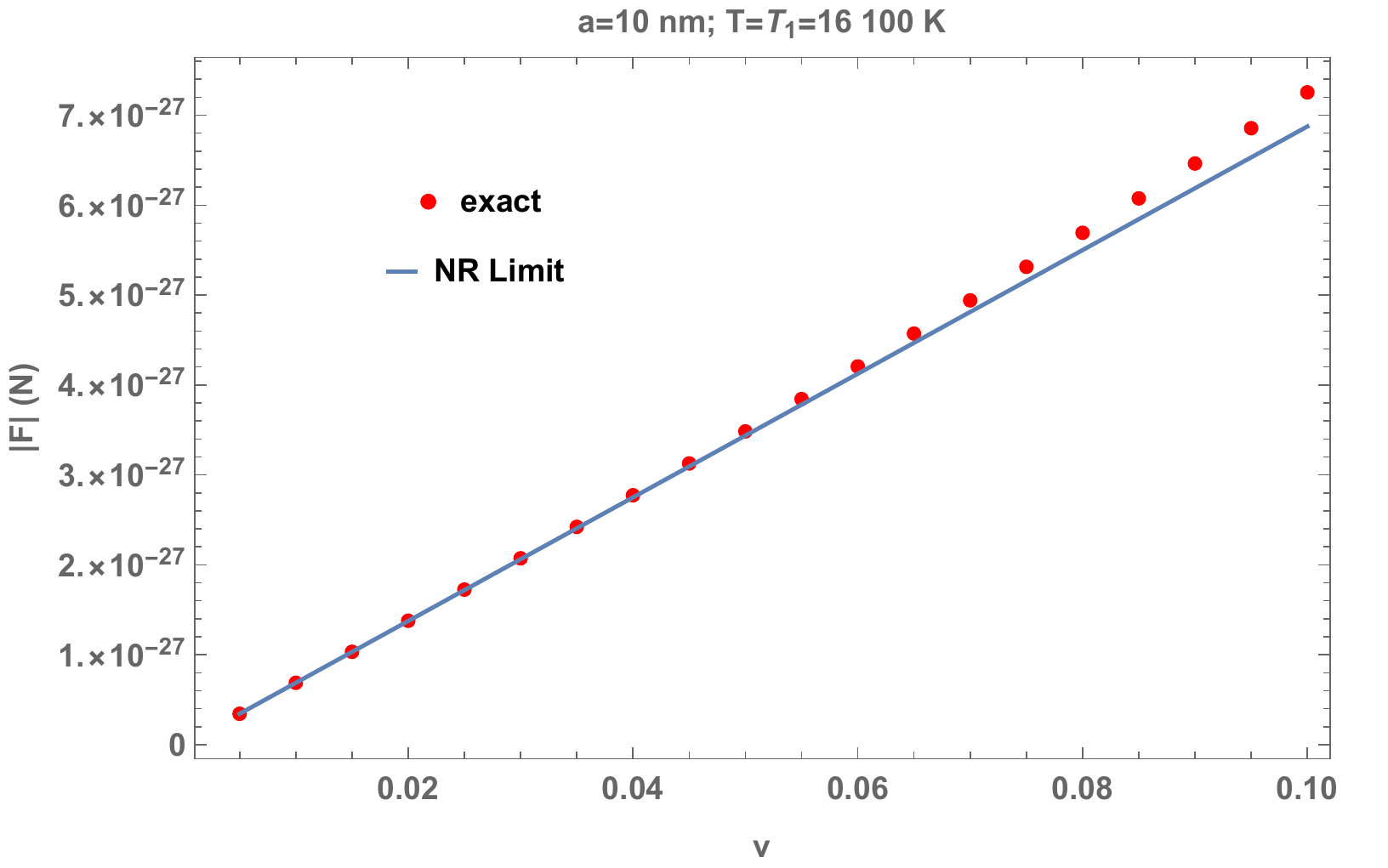}%
}
\subfloat[]{\label{FvR}%
\includegraphics[width=0.48\linewidth]{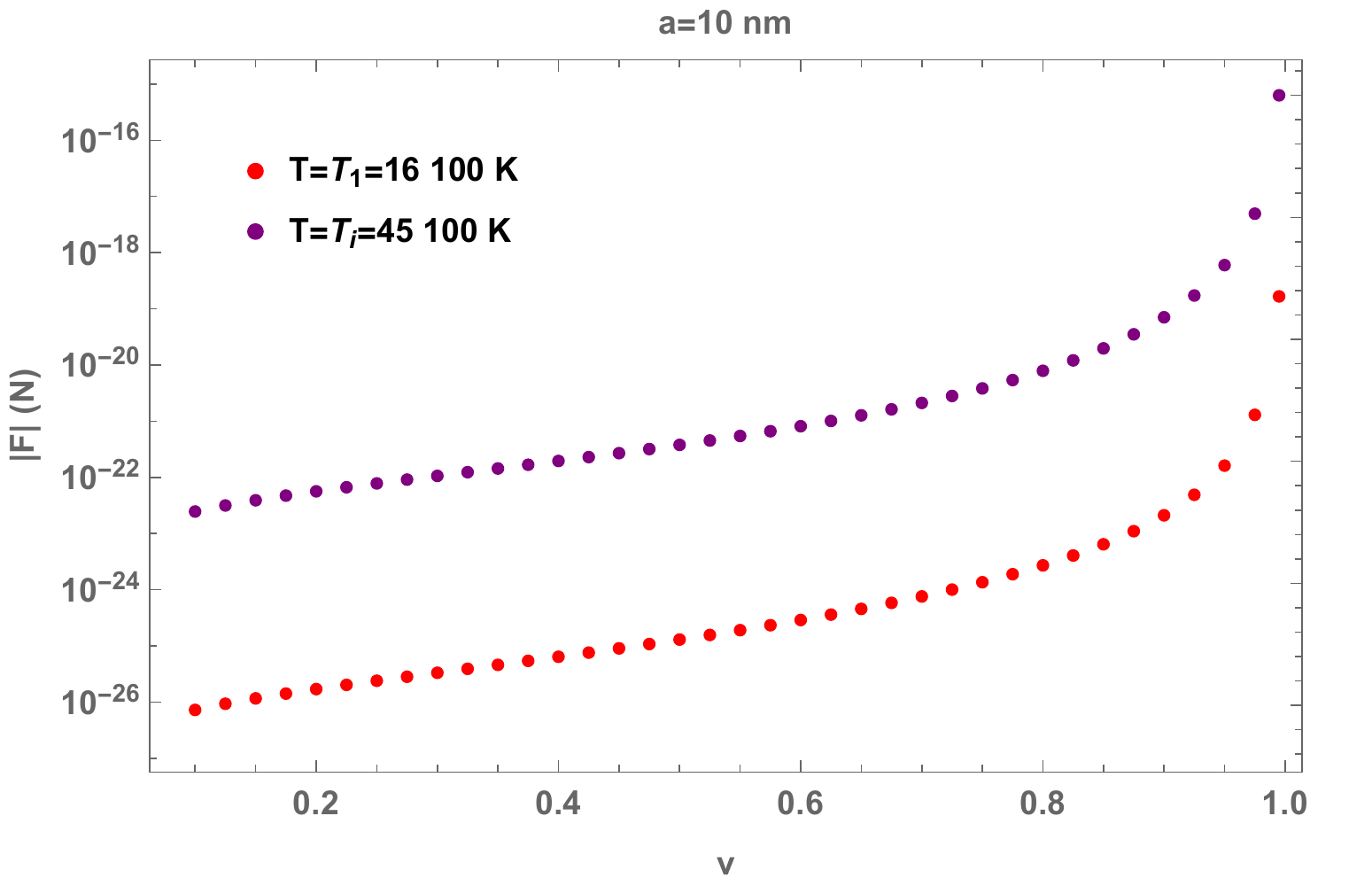}%
}
\caption{The velocity dependence of the magnitude of the total frictional force on a cesium atom at a distance of $ a=10\, \rm{nm} $ away from the PC plate. (a) At the first excitation temperature, $ T_{1}=16\,100\, \rm{K} $, the frictional force is plotted as a function of velocity for velocities, $ v\in [0.005,0.100] $. The red dots show the exact numerical results based on Eq.~\eqref{scale} and Eq.~\eqref{main}. The blue solid line shows the term linear in $ v $ obtained using the nonrelativistic approximation detailed in Appendix  \ref{NR}. (b) In the more relativistic regime, $ v\in [0.100,0.995] $, the red dots show the total frictional force at the first excitation temperature, $ T_{1}=16\,100\, \rm{K}  $, while the purple dots show the numerical results extrapolated to the ionization temperature, $ T_{i}=45\,100\, \rm{K} $. For the maximum velocity shown in the figure, $ v=0.995 $, the magnitude of the total friction is  $ 1.66\cross 10^{-19} \,\rm{N} $ at $ T_{1} $ and $ 6.30\cross 10^{-16}\,\rm{N} $ at $ T_{i} $.}
\label{Fv}
\end{figure*}

Finally, QFPC depends on the velocity of the atom. As is shown in Fig.~\ref{FvNR}, the magnitude of the frictional force is linear in $ v $ for very small velocities; however, the velocity dependence becomes more prominent for larger velocities. In Fig.~\ref{FvR}, we not only plot the total frictional force at the first excitation temperature, $ T_{1}=16\,100\,\rm{K} $, but also extrapolate our numerical results to the ionization temperature of the cesium atom, $ T_{i}=45\,100\,\rm{K} $ \cite{Weber:csenergy}. Above $ T_{i} $, the outermost electron will be stripped off the atom so that the cesium atom cannot stay neutral. It is therefore not feasible experimentally to detect the quantum friction on an atom above its ionization temperature. In between $ T_{1} $ and $ T_{i} $, the atom can be excited, though not ionized. Now, the frequencies corresponding to the transition of the atom's internal energy levels become important in evaluating QFPC. At these frequencies, the polarizability of the atom develops an imaginary part \cite{Lach:bbf}, which results in a QFPC that is first order in the polarizability. This effect is not included in the results we show for $ T=T_{i} $. In addition, by employing the static value for the polarizability, we underestimate the magnitude of the second order QFPC,  because atoms in excited states, e.g., Rydberg atoms\footnote{Even though Rydberg atoms possess much larger polarizabilities, which presumably will enhance the resulting frictional effect, we are unsure whether they could be appropriate candidates for experimental consideration, because blackbody radiation induces transitions to lower $ n $ states and reduces the lifetime of the Rydberg states.  Even at room temperature, transitions induced by blackbody radiation can contribute more to the decay rate than the spontaneous transitions \cite{Beterov:Rydberg}. At higher temperatures, the  transition rate induced by blackbody radiation is even larger.}, tend to have much larger polarizabilities.

\section{CONCLUSIONS AND OUTLOOK}
In this paper, we calculate the frictional force induced by fluctuations of the electromagnetic field on a neutral, nondissipative atom moving parallel to a perfectly conducting (PC) plate, which we term QFPC for brevity. This friction exists in second order in the polarizability of the atom and reduces to the quantum vacuum friction previously explored \cite{Xin:eqf1} in the limit of large distance from the plate or high temperature. At short distances or low temperatures, however, the PC plate modifies the behavior of the frictional force. For an isotropic atom, the frictional force is found to be negative definite (a drag) and twice the magnitude of the quantum vacuum friction felt by the same atom moving through blackbody radiation without a PC plate. Interestingly, the contribution to QFPC from a particular polarization state of the atom is positive (a push). However, this contribution turns out to be always subdominant to the negative contributions from the other polarization states. As a result, the total transverse force on the moving atom remains a drag. The magnitude of the friction seems to be too tiny to be observed unless the atom is made to move in a very hot background, at very high velocities. It is then fitting to make some comments about the experimental conditions on temperature and velocity.

First, the temperature is absolutely bounded by the ionization temperature of the atom, $ T_{i} $, because the atom will no longer stay neutral above $ T_{i} $. In this paper, we have used a static polarizability for the atom in the numerical calculations, which can only be justified if the atom remains in its ground state. The temperature is therefore further bounded by the temperature, $ T_{1} $, corresponding to the first excitation energy of the atom. In principle, however, one can calculate QFPC on the atom up to its ionization temperature if its dynamical polarizability is known for a sufficiently wide frequency spectrum. 

Of course, for experiments, the material which approximates a PC plate is very likely to give a more restrictive bound on the temperature. For example, the standard candidate, gold, will melt at $ 1337\,\rm{K} $. Even if one could imagine using liquid metals to mimic the perfectly conducting plate, the temperature is still bounded by the boiling point of the metal. Tungsten has the highest boiling point among metals, $ 6203\,\rm{K} $, which is still much lower than the typical ionization temperature, $ T_{i} $, of an atom. (For the cesium atom discussed in the paper, $ T_{i}=45\,100\,\rm{K} $.) This reality might motivate us to study QFPC for situations when the plate is not in thermal equilibrium with the radiation background. For the Casimir-Polder force (the force normal to the plate), such a scenario has been studied both theoretically \cite{Antezza:New}  and experimentally \cite{Obrecht:CP}.

Another apparent challenge to any feasible experiment is accelerating neutral atoms to relativistic velocities. But, in fact, it is possible nowadays to manipulate the conventional ion accelerators so that fast ions can be converted to neutral atoms with little change in momentum. For example, in Ref.~\cite{Dalui:accelerate}, the maximum kinetic energy obtained for a copper atom is $ 1\,\rm{MeV} $, which is equivalent to a velocity of $ 0.0058 $, after conversion using the relativistic formula for kinetic energy, $ K=(\gamma-1)m $.

This paper only considers a very idealized background with the PC plate. For a surface with a real finite index of refraction, $ n $, there will be induced Cherenkov friction \cite{Pieplow:Cherenkov} on the moving particle if it moves at a velocity above the Cherenkov threshold, $ v>1/n $. If one further allows the surface to have dissipation, which is unavoidable in reality and perhaps induces an even greater frictional effect, the problem becomes complicated by the presence of several different mechanisms that give rise to friction. Ref.~\cite{Reiche:Wading} provides a recent overview of this complicated subject with many useful references; however, it mainly focuses on only zero-temperature effects. Other works, like Ref.~\cite{Oelschlager:viscosity}, do include finite-temperature effects but the discussion is restricted to only the nonrelativistic regime. In the future, we intend to calculate the quantum friction associated with a dispersive and dissipative surface fully for arbitrary temperatures and relativistic velocities. Alternatively, one could still assume a perfectly conducting boundary but allow the moving particle itself to be intrinsically dissipative. The resulting QFPC will then be a modification of the quantum vacuum friction studied in Ref.~\cite{Xin:eqf2}, where an independent temperature of the particle comes into play. We will discuss such QFPC in a subsequent paper. 

%acknowledgements
\begin{acknowledgments}
We thank the US National Science Foundation, grant No. 2008417, for partial support of this work. We thank Stephen Fulling, Prachi Parashar, Shadi Rezaei, Dylan Michael DelCol, and Venkat Abhignan for many helpful discussions. We thank Matthias Kr\"{u}ger for pointing us to the analogy of the colloid-wall interaction in hydrodynamics. This paper reflects solely the authors' personal opinions and does not represent the opinions of the authors' employers, present and past, in any way.
\end{acknowledgments}

%appendix
\appendix
\section{THE TRANSFORMATION OF THE REDUCED GREEN'S DYADIC}\label{gprime}
In this appendix, we provide the connection between the reduced Green's dyadic in frame $ \mathcal{P} $, $ \bm{g}' $, and that in frame $ \mathcal{R} $, $ \bm{g} $. It is straightforwardly obtained by considering the Lorentz transformation of the electromagnetic field and applying the FDT in both frames consistently.   Note the transformation of the material properties like $ \bm{\varepsilon} $ or $ \bm{\mu} $ is never invoked because we eventually express the quantum friction in terms of $ \bm{g} $ instead of $ \bm{g}' $. 

In writing down the connection between $ \bm{g} $ and $ \bm{g}' $, we will use $ (\omega', \bm{k}'_{\perp}) $ for the frequency and momentum in the atom's rest frame ($ \mathcal{P} $), and $(\omega, \bm{k}_{\perp} ) $ for those transformed into the rest frame of the radiation ($ \mathcal{R} $), 
\begin{equation}\label{wk}
\omega=\gamma (\omega'+k'_{x}v), \qquad k_{x}=\gamma (k'_{x}+\omega'v), \qquad k_{y}=k_{y}'.
\end{equation}

Below all components of $ \bm{g}' $ are expressed in terms of components of $ \bm{g} $:

\begin{align}
g_{xx}'(\omega', \bm{k}'_{\perp};z,\tilde{z})&=g_{xx}(\omega,\bm{k}_{\perp};z,\tilde{z}),\nonumber\\
g_{yy}'(\omega', \bm{k}'_{\perp};z,\tilde{z})&=\frac{1}{(\omega'+k_{x}'v)^{2}}\left(\frac{{\omega'}^{2}}{\gamma^{2}}g_{yy}+{k_{y}'}^{2}v^{2}g_{xx}+\frac{\omega'}{\gamma}k_{y}'v g_{xy}+\frac{\omega'}{\gamma}k_{y}'v g_{yx}\right)(\omega,\bm{k}_{\perp};z,\tilde{z}),\nonumber\\
g_{zz}'(\omega', \bm{k}'_{\perp};z,\tilde{z})&=\frac{1}{(\omega'+k_{x}'v)^{2}}\left(\frac{{\omega'}^{2}}{\gamma^{2}}g_{zz}+v^{2}\partial_{z}\partial_{\tilde{z}} g_{xx}+i\frac{\omega'}{\gamma}v\partial_{\tilde{z}}g_{zx}-i\frac{\omega'}{\gamma}v\partial_{z}g_{xz}\right)(\omega,\bm{k}_{\perp};z,\tilde{z}),\nonumber\\
g_{xy}'(\omega', \bm{k}'_{\perp};z,\tilde{z})&=\frac{1}{\omega'+k_{x}'v} \left(\frac{\omega'}{\gamma}g_{xy}+k_{y}'v g_{xx}\right)(\omega,\bm{k}_{\perp};z,\tilde{z}),\nonumber\\
g_{yx}'(\omega', \bm{k}'_{\perp};z,\tilde{z})&=\frac{1}{\omega'+k_{x}'v} \left(\frac{\omega'}{\gamma}g_{yx}+k_{y}'v g_{xx}\right)(\omega,\bm{k}_{\perp};z,\tilde{z}),\nonumber\\
g_{zx}'(\omega', \bm{k}'_{\perp};z,\tilde{z})&=\frac{1}{\omega'+k_{x}'v} \left(\frac{\omega'}{\gamma}g_{zx}+iv\partial_{z} g_{xx}\right)(\omega,\bm{k}_{\perp};z,\tilde{z}),\nonumber\\
g_{xz}'(\omega', \bm{k}'_{\perp};z,\tilde{z})&=\frac{1}{\omega'+k_{x}'v} \left(\frac{\omega'}{\gamma}g_{xz}-iv\partial_{\tilde{z}} g_{xx}\right)(\omega,\bm{k}_{\perp};z,\tilde{z}),\nonumber\\
g_{yz}'(\omega', \bm{k}'_{\perp};z,\tilde{z})&=\frac{1}{(\omega'+k_{x}'v)^{2}}\left(\frac{{\omega'}^{2}}{\gamma^{2}}g_{yz}-ik_{y}'v^{2}\partial_{\tilde{z}}g_{xx}+i\frac{\omega'}{\gamma}v\partial_{\tilde{z}} g_{yx}-\frac{\omega'}{\gamma}k_{y}'v g_{xz}\right)(\omega,\bm{k}_{\perp};z,\tilde{z}),\nonumber\\
g_{zy}'(\omega', \bm{k}'_{\perp};z,\tilde{z})&=\frac{1}{(\omega'+k_{x}'v)^{2}}\left(\frac{{\omega'}^{2}}{\gamma^{2}}g_{zy}+ik_{y}'v^{2}\partial_{z}g_{xx}-i\frac{\omega'}{\gamma}v\partial_{z} g_{xy}-\frac{\omega'}{\gamma}k_{y}'v g_{zx}\right)(\omega,\bm{k}_{\perp};z,\tilde{z}).
\end{align}

\section{THE FORM OF THE REDUCED GREEN'S DYADIC}\label{g}
In this appendix, we give the explicit form of the reduced Green's dyadic used in the paper.

The Green's dyadic $\bm{\Gamma}(\bm{r},\tilde{\bm{r}};\omega)$ in frequency space satisfies the following differential equation,
\begin{equation}\label{eqB-1}
\left[ - \bm{\varepsilon}(\bm{r};\omega)+\frac{1}{\omega^{2}}\curl\bm{\mu}^{-1}(\bm{r};\omega)\cdot\curl\right]\bm{\Gamma}(\bm{r},\tilde{\bm{r}};\omega)=\bm{1}\delta(\bm{r}-\tilde{\bm{r}}).
\end{equation}
where $ \bm{\varepsilon}(\bm{r};\omega) $ and $ \bm{\mu}(\bm{r};\omega) $ are the permittivity and permeability at the field point $ \bm{r} $. In deriving Eq.~\eqref{eqB-1}, we have ignored the spatial dispersion effects so that these susceptibilities are local in space. The geometry of the problem we consider possesses translational symmetry in the $ x$-$y $ plane, which permits us to Fourier transform the Green's dyadic in these spatial directions,
\begin{equation}\label{eqB-2}
\bm{\Gamma}(\bm{r},\tilde{\bm{r}};\omega)=\int \frac{d^{2}\bm{k}_{\perp}}{(2\pi)^{2}} e^{i\bm{k}_{\perp}\cdot (\bm{r}_{\perp}-\tilde{\bm{r}}_{\perp})} \bm{g}(z,\tilde{z};\omega,\bm{k}_{\perp}).
\end{equation}

In this paper, we always evaluate the Green's dyadic at the position of the particle, where the permittivity and the permeability become scalars and take the vacuum value, $ \varepsilon=\mu=1 $. The reduced Green's dyadic $ \bm{g} $ then takes the special form
\begin{align}
\bm{g}(z,\tilde{z};\omega,\bm{k}_{\perp})=&\mqty(\frac{k_{x}^{2}}{k^{2}}\partial_{z}\partial_{\tilde{z}}g^{H}+\frac{k_{y}^{2}}{k^{2}}\omega^{2}g^{E}
&
\frac{k_{x}k_{y}}{k^{2}}\partial_{z}\partial_{\tilde{z}}g^{H}-\frac{k_{x}k_{y}}{k^{2}}\omega^{2}g^{E}
&
\quad ik_{x}\partial_{z}g^{H}
\\\\
\frac{k_{x}k_{y}}{k^{2}}\partial_{z}\partial_{\tilde{z}}g^{H}-\frac{k_{x}k_{y}}{k^{2}}\omega^{2}g^{E}
&
\frac{k_{y}^{2}}{k^{2}}\partial_{z}\partial_{\tilde{z}}g^{H}+\frac{k_{x}^{2}}{k^{2}}\omega^{2}g^{E}
&
\quad ik_{y}\partial_{z}g^{H}
\\\\
-ik_{x}\partial_{\tilde{z}}g^{H}
&
-ik_{y}\partial_{\tilde{z}}g^{H}
&
\quad k^{2}g^{H}).
\end{align}
The scalar Green's functions that construct the Green's dyadic consist of a bulk part and a scattering part,
\begin{equation}\label{gEH}
g^{E,H}(z,\tilde{z};\omega,k)=\frac{1}{2\kappa}e^{-\kappa|z-\tilde{z}|}+\frac{r^{E,H}}{2\kappa}e^{-\kappa(z+\tilde{z})},
\end{equation}
with the reflection coefficients
\begin{equation}
r^{E}=\frac{\kappa-\kappa'/\mu}{\kappa+\kappa'/\mu}, \qquad
r^{H}=\frac{\kappa-\kappa'/\varepsilon}{\kappa+\kappa'/\varepsilon}.
\end{equation}
Here, $ \varepsilon $ and $ \mu $ are the permittivity and permeability of the reflecting surface, which is assumed to be homogeneous and isotropic for simplicity, and $ \kappa $ and $ \kappa' $ are the propagation wave numbers associated with the vacuum and the surface, respectively, given by
\begin{equation}
\kappa^2=k^{2}-\omega^{2},\qquad
\kappa'^2=k^{2}-\omega^{2}\varepsilon\mu.
\end{equation}
In certain regions for $ \omega $ and $ \bm{k}_{\perp} $, these wave numbers could develop an imaginary part, which is crucial for discussions of dissipative forces like quantum friction. In those regions, the branch is so chosen that the retarded requirement of the Green's dyadic is guaranteed, 
\begin{equation}\label{branch}
\kappa\to -i\sgn(\omega)\sqrt{\omega^{2}-k^{2}}, \quad \omega^{2}>k^2; \qquad \kappa'\to -i\sgn(\omega)\sqrt{\omega^{2}\varepsilon\mu-k^{2}}, \quad \omega^{2}\varepsilon\mu>k^2.
\end{equation}
Note that $ \kappa $  becomes odd in $ \omega $ in the region where it develops imaginary part.

In the perfectly (electrically) conducting limit for the  surface considered in this paper, the permittivity and permeability take the extreme values \cite{Sihvola:PEC}, 
\begin{equation}\label{PEC}
\varepsilon\to\infty, \qquad \mu\to 0
\end{equation}
so that the reflection coefficients simplify to be
\begin{equation}\label{rPC}
r^{E,H}=\mp 1.
\end{equation}

\section{THE ABSENCE OF ZERO TEMPERATURE QUANTUM FRICTION IN THE PRESENCE OF A DIAPHANOUS MEDIUM}
In this appendix, we supply a proof for why no zero temperature QF should arise for the vacuum case and the PC case. Further, we extend the claim to include any diaphanous, nondissipative medium with the property $ \varepsilon\mu=1 $.

The general QF for an atom, Eq.~\eqref{starting} can be rewritten as the following when the temperature is set to be zero: 
\begin{equation}\label{zero}
F=2\int_{0}^{\infty}\frac{d\omega}{2\pi}\int\frac{d^{2}\bm{k}_{\perp}}{(2\pi)^2}\frac{d^{2}\bar{\bm{k}}_{\perp}}{(2\pi)^2} \bar{k}_{x} \tr \left[\bm{\alpha}(\omega)\cdot \imaginary\, \bm{g}'(\omega,\bm{k}_{\perp};a,a)\cdot\bm{\alpha}(\omega)\cdot \imaginary\, \bm{g}'(\omega,\bar{\bm{k}}_{\perp};a,a)\right]\left[\sgn(\omega+\bar{k}_{x}v)-\sgn(\omega+k_{x}v)\right].
\end{equation}
To obtain Eq.~\eqref{zero}, we have exchanged $ k_{x} $ and $ \bar{k}_{x} $ for the second term in Eq.~\eqref{starting} and used the evenness of the integrand under the total reflection of its frequency and wave vector arguments $ (\omega, \bm{k}_{\perp}, \bar{\bm{k}}_{\perp}) \to (-\omega, -\bm{k}_{\perp}, -\bar{\bm{k}}_{\perp}) $. In order to make the argument clearer, let us change the $ \bm{k}_{\perp} $ and $ \bar{\bm{k}}_{\perp} $ into dimensionless variables using $ \omega $ as a positive scale, 
\begin{equation}\label{scaling}
k_{x}=\omega x, \quad k_{y}=\omega y, \quad \bar{k}_{x}=\omega \bar{x}, \quad \bar{k}_{y}=\omega \bar{y}.
\end{equation}
The frictional force now reads
\begin{equation}\label{Fsca}
F=\frac{1}{16\pi^5}\int_{0}^{\infty}d\omega \,\omega^5 \int dx dy d\bar{x} d\bar{y} \ \bar{x} \tr \left[\bm{\alpha}(\omega)\cdot \imaginary\, \bm{g}'(\omega, \omega x, \omega y)\cdot\bm{\alpha}(\omega)\cdot \imaginary\, \bm{g}'(\omega, \omega \bar{x}, \omega \bar{y})\right]\left[\sgn(1+\bar{x}v)-\sgn(1+xv)\right],
\end{equation}
where we have suppressed the spatial $ z $ coordinates of the Green's dyadics.
The difference in the $ \sgn $ functions can be translated into limits for the $ x $ and $ \bar{x} $ integrals, leading to
\begin{equation}\label{translation}
F=\frac{1}{8\pi^5}\int_{0}^{\infty}d\omega \,\omega^5 \int dy d\bar{y} \left[\int_{-\infty}^{-\frac{1}{v}} dx \int_{-\frac{1}{v}}^{\infty} d\bar{x}-\int_{-\frac{1}{v}}^{\infty} dx \int_{-\infty}^{-\frac{1}{v}} d\bar{x}\right]\,\bar{x} \tr\left[ \bm{\alpha}(\omega)\cdot \imaginary\, \bm{g}'(\omega, \omega x, \omega y)\cdot\bm{\alpha}(\omega)\cdot \imaginary\, \bm{g}'(\omega, \omega \bar{x}, \omega \bar{y})\right].
\end{equation}
By exchanging $ x $ and $ \bar{x} $ again for the second term inside the bracket of Eq.~\eqref{translation}, we find the frictional force becomes
\begin{equation}\label{back}
F=\frac{1}{8\pi^5}\int_{0}^{\infty}d\omega \,\omega^5 \int dy d\bar{y} \int_{-\infty}^{-\frac{1}{v}} dx \int_{-\frac{1}{v}}^{\infty} d\bar{x}\,(\bar{x}-x) \tr \left[\bm{\alpha}(\omega)\cdot \imaginary\, \bm{g}'(\omega, \omega x, \omega y)\cdot\bm{\alpha}(\omega)\cdot \imaginary\, \bm{g}'(\omega, \omega \bar{x}, \omega \bar{y})\right].
\end{equation}
Now, the limit on $ x $ prevents the vacuum propagation wave number of the first reduced Green's dyadic, $ \kappa $, from developing an imaginary part, because of 
\begin{equation}\label{realdefinite}
\kappa^2=k^2-\omega^2=\omega^{2}(x^2+y^2-1)>0, \qquad x<-\frac{1}{v}.
\end{equation}

For the simplest vacuum situation where only the diagonal components of the Green's dyadic contribute to the integral (see Appendix A of Ref.~\cite{Xin:eqf1} for a detailed discussion), the anti-Hermitian part reduces to the ordinary imaginary part. But the only possible source of an imaginary part for the first Green's dyadic in Eq.~\eqref{back}, $ \kappa $, is now real definite. As a result, the zero temperature QVF vanishes.

For backgrounds other than vacuum, zero temperature quantum friction exists in general because the propagation wave number associated with the medium can become imaginary since
\begin{equation}\label{indefinite}
\kappa'^2=k^2-\omega^2\varepsilon\mu=\omega^{2}(x^2+y^2-\varepsilon\mu)
\end{equation}
does not have a definite sign. A diaphanous medium with the special property,
\begin{equation}\label{diaphanous}
\varepsilon\mu=1,
\end{equation}
however, is an exception, for which the propagation wave number coincides with the vacuum one, $ \kappa'=\kappa $. This nice coincidence renders the reflection coefficients to be real definite as long as $ \varepsilon $ and $ \mu $ are real,
\begin{equation}
r^E=\frac{\mu-1}{\mu+1}=\frac{1-\varepsilon}{1+\varepsilon},\qquad r^H=\frac{\varepsilon-1}{\varepsilon+1}.
\end{equation}
Therefore, the only source of the imaginary part in the scalar Green's functions Eq.~\eqref{gEH} is still the $ \kappa $ as in the vacuum case. It can be further checked that the anti-Hermitian part of $ \bm{g}' $ vanishes unless $ \kappa $ develops an imaginary part even though the off-diagonal components of the Green's dyadic and the transformation between $ \bm{g}' $ and $ \bm{g} $ needs to be taken into account. Again, recalling Eq.~\eqref{realdefinite}, the zero temperature QF must be absent even if such a diaphanous medium is present in the background. 

Now, apparently, both the perfect conductor defined by Eq.~\eqref{PEC} and Eq.~\eqref{rPC} and the vacuum background can be deemed as members of the family of diaphanous materials, for which the total reflection coefficient $ r^E+r^H=0 $.

\section{THE NONRELATIVISTIC LIMIT OF QFPC}\label{NR}
In this appendix, we obtain the nonrelativistic (NR) limit of QFPC  directly from the expressions in Eq.~\eqref{main} and Eq.~\eqref{scale}, where we have already replaced the dynamical polarizability with the static polarizability. We will use $ F^{\rm{XZ}} $ in particular as an example to illustrate the procedure:
\begin{align}
\label{uXZ}
F^{\rm{XZ}}=&\frac{\alpha_{xx}(0)\alpha_{zz}(0)}{8\pi^{3}(2a)^8}\int_{0}^{\infty} \,dx \, x^{3}\left[3x\cos x+(x^{2}-3)\sin x\right]\nonumber\\
\cross & \int_{-1}^{1} du \,u \sqrt{1-u^2}\,J_{1}\left(x\sqrt{1-u^2}\right)\frac{1}{e^{xz\gamma(1+uv)}-1}.
\end{align}
In the NR limit, the exponential factor can be expanded in $ v $. Keeping only up to the term linear in $ v $, we obtain
\begin{align}
\label{linear}
F^{\rm{XZ}}=&\frac{\alpha_{xx}(0)\alpha_{zz}(0)}{8\pi^{3}(2a)^8}\int_{0}^{\infty} \,dx \,x^{3}\left[3x\cos x+(x^{2}-3)\sin x\right]\nonumber\\
\cross & \int_{-1}^{1} du \,u \sqrt{1-u^2}\,J_{1}\left(x\sqrt{1-u^2}\right)\left[\frac{1}{e^{xz}-1}- uv \frac{xze^{xz}}{(e^{xz}-1)^2}\right].
\end{align}
Note the term constant in $ v $ vanishes because of its oddness in $ u $, as there should be no spontaneous quantum propulsion for a reciprocal point particle. See Ref.~\cite{Kim:torque}. We are left with the term linear in $ v $ as expected. The $ u $ integral can then be easily carried out and we obtain
\begin{align}
\label{vz}
F^{\rm{XZ}}&=\frac{\alpha_{xx}(0)\alpha_{zz}(0)}{16\pi^{3}(2a)^8}v I(z),\qquad I(z)=z\int_{0}^{\infty} dx \frac{\left[3x\cos x +(x^2-3)\sin x\right]^2}{\sinh^{2}(xz/2)}.
\end{align}

Now we focus on $ I(z) $, which carries all the $ z $ dependence of $ F^{\rm{XZ}} $, and rewrite it as
\begin{equation}
I(z)=-4\int_0^{\infty}dx\,\left[3x\cos x +\left(x^2-3\right)\sin x\right]^2 \frac{d}{dx}\left(\frac{1}{e^{xz}-1}\right).
\end{equation}
Using integration by parts, this becomes
\begin{equation}
\begin{aligned}
I(z)&=4\int_0^{\infty}dx\,\left[x\left(2x^2+3\right)+x^2\left(x^2-6\right)\sin (2x)+x\left(4x^2-3\right)\cos(2x)\right]\frac{1}{e^{xz}-1}\\
&=4\left\{\int_0^{\infty}dx\,\frac{\left(2x^3+3x\right)}{e^{xz}-1}+\left.\left[\frac{d^4}{db^4}-4\frac{d^3}{db^3}+6\frac{d^2}{db^2}-3\frac{d}{db}\right]\int_0^{\infty}dx\,\frac{\sin(bx)}{e^{xz}-1}\right\}\right|_{b=2}\\
&=4\left\{\int_0^{\infty}dx\,\frac{\left(2x^3+3x\right)}{e^{xz}-1}+\left.\left[\frac{d^4}{db^4}-4\frac{d^3}{db^3}+6\frac{d^2}{db^2}-3\frac{d}{db}\right]\left[\frac{\pi}{2z}\coth\left(\frac{b\pi}{z}\right)-\frac{1}{2b}\right]\right\}\right|_{b=2}\\
%&=\left\{\frac{8\pi^4}{15z^4}+\frac{2\pi^2}{z^2}-9+\left.\frac{2\pi}{z}\left[\frac{d^4}{db^4}-4\frac{d^3}{db^3}+6\frac{d^2}{db^2}-3\frac{d}{db}\right]\coth\left(\frac{b\pi}{z}\right)\right\}\right|_{b=2}\\
%&=\frac{8\pi^4}{15z^4}+\frac{2\pi^2}{z^2}-9-\frac{2\pi^2}{z^2}\left.\left[\frac{d^3}{db^3}-4\frac{d^2}{db^2}+6\frac{d}{db}-3\right]\operatorname{csch}^2\left(\frac{b\pi}{z}\right)\right|_{b=2}\\
&=\frac{8\pi^4}{15z^4}+\frac{2\pi^2}{z^2}-9\\
&\quad+\left\lbrace\frac{16\pi^5}{z^5}\!\left[3\coth^2\left(\frac{2\pi}{z}\right)\!-\!2\right]\!\coth\left(\frac{2\pi}{z}\right)+\frac{16\pi^4}{z^4}\!\left[3\coth^2\left(\frac{2\pi}{z}\right)\!-\!1\right]+\frac{24\pi^3}{z^3}\!\coth\left(\frac{2\pi}{z}\right)\!+\frac{6\pi^2}{z^2}\right\rbrace\!\operatorname{csch}^2\left(\frac{2\pi}{z}\right).
\end{aligned}
\end{equation}
For $ z\ll 1 $, the leading term of $ I(z) $ is $ 8\pi^{4}/15z^{4} $. In appendix \ref{limit}, we will show both analytically and numerically that $ F^{\rm{XZ}} $ behaves as $ 1/z^{4} $ in the small $ z $ limit, even for relativistic velocities.  When $ I(z) $ is expanded for $ z\gg 1 $, on the other hand, multiple cancellation occurs and the leading term of $ I(z) $ is found to be $ 1024\pi^{10}/1485z^{10} $. This agrees with the large $ z $ limit for arbitrary velocities already obtained in Eq.~\eqref{key}.

We have also found the other contributions to the QFPC all have a nonvanishing term linear in $ v $. The procedure outlined in this Appendix works to extract the correct NR limits of these other contributions to QFPC as well.

\section{SMALL AND LARGE z LIMITS OF QFPC}\label{limit}
In this appendix, we show how to obtain the small and large $z$ limits of QFPC for all $v$. We will, again, focus on the $\rm{XZ}$ polarization contribution.

Rewriting Eq.~\eqref{mainXZ} as
\begin{equation}\label{E1}
\mathcal F^{\rm{XZ}}(x, v, z)=-\sqrt{\pi}\,2^{\frac32}x^{-\frac32}J_{\frac52}(x) \int_{-1}^1 du\, u\sqrt{1-u^2}\, J_1\left(x\sqrt{1-u^2}\right)\frac{1}{e^{x\gamma z(1+uv)}-1}
\end{equation}
and explicitly expanding the thermal occupation factor as a Maclaurin series in the $v$ variable, but retaining the implicit dependence of $\gamma$ on $v$, 
\begin{equation}\label{E2}
\frac{1}{e^{x\gamma z(1+uv)}-1}=\sum_{n=0}^{\infty}\frac{v^n}{n!}\left. \left[\frac{\partial^n}{\partial v^n} \frac{1}{e^{x\gamma z(1+uv)}-1}\right]\right|_{v=0}
=\sum_{n=0}^{\infty}\frac{v^nu^n}{n!}(x\gamma z)^n \frac{\partial^n}{\partial (x\gamma z)^n} \frac{1}{e^{x\gamma z}-1},
\end{equation}
we obtain
\begin{equation}\label{E3}
\begin{aligned}
\mathcal F^{\rm{XZ}}(x, v, z)&=-\sqrt{\pi}\,2^{\frac32}x^{-\frac32}J_{\frac52}(x)\sum_{n=0}^{\infty}\frac{v^n}{n!}\int_{-1}^1du\,u^{n+1}\sqrt{1-u^2} \,J_1\left(x\sqrt{1-u^2}\right) z^n \frac{\partial^n}{\partial z^n}\frac{1}{e^{x\gamma z}-1}\\
&=-\pi J_{\frac52}(x) \sum_{m=0}^{\infty}\frac{v^{2m+1}}{m!}2^{2-m}x^{-(m+3)}J_{m+\frac52}(x) \,z^{2m+1}\frac{\partial^{2m+1}}{\partial z^{2m+1}}\frac{1}{e^{x\gamma z}-1},
\end{aligned}
\end{equation}
where we have noticed that the odd $ n $ terms vanishes because of the symmetry of the integrand for the $ u $ integral.

It follows from Eq.~\eqref{scale} and Eq.~\eqref{E3} that
\begin{equation}\label{E4}
f^{XZ}(v, z)= -\pi \sum_{m=0}^{\infty}\frac{v^{2m+1}}{m!}2^{2-m}z^{2m+1}\frac{\partial^{2m+1}}{\partial z^{2m+1}}\int_0^{\infty}dx\, x^{4-m}J_{\frac52}(x) \,J_{m+\frac52}(x)\,\frac{1}{e^{x\gamma z}-1},
\end{equation}
which may be cast in forms suitable for small or large $z$ by employing representations of the integrand (other than the thermal occupation factor) that are appropriate for large or small $x$, respectively.

\begin{figure}
 \includegraphics[width=0.70\linewidth]{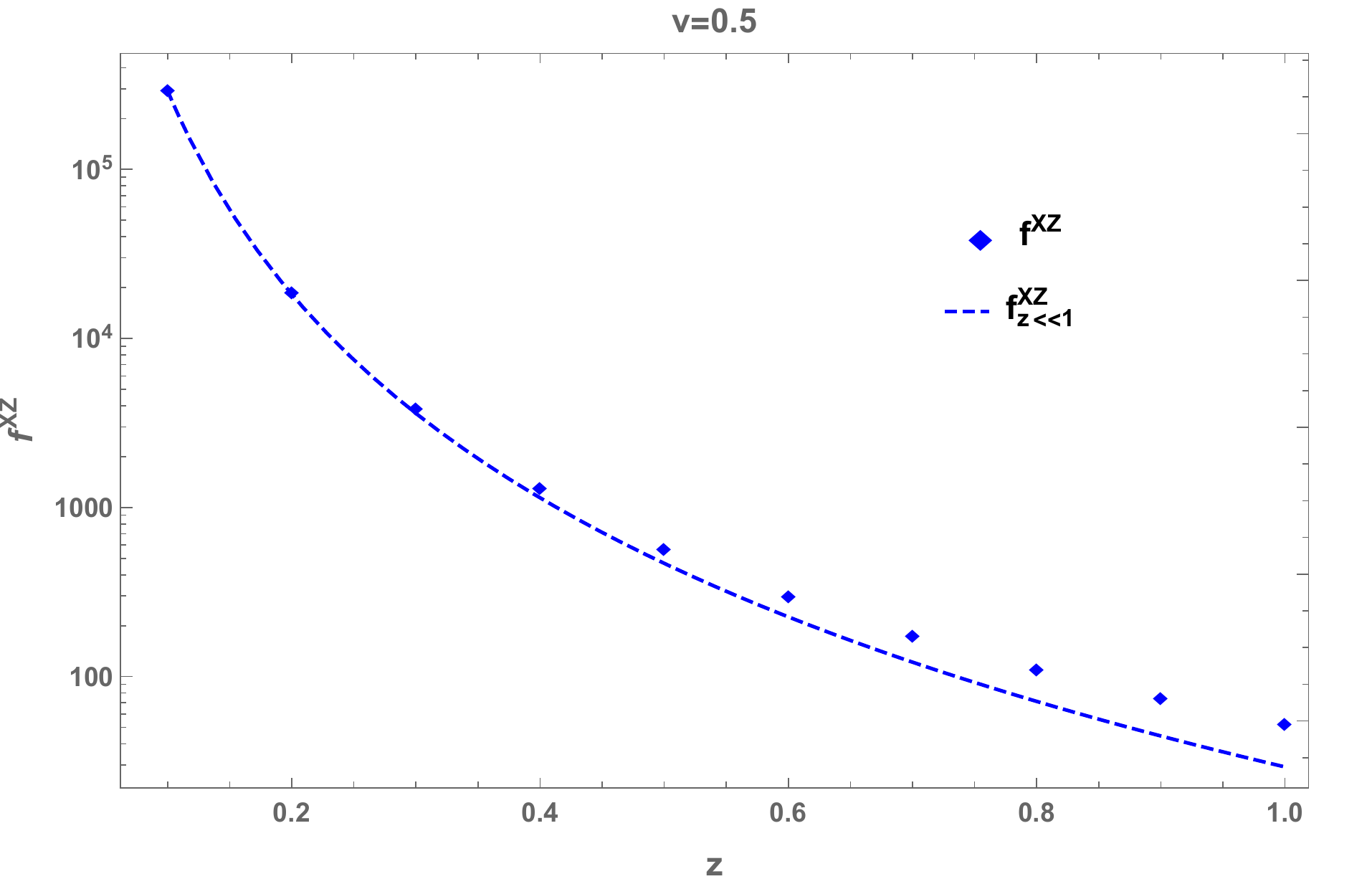}
 \caption{At fixed velocity $ v=0.5 $, the numerical results for $ f^{\rm{XZ}} $ (dots) and its small $ z $ approximation (dashed line) obtained in Eq.~\eqref{E7} are shown for $ z\in [0,1] $.} \label{FigE}
 \end{figure}

Thus, the finite series representation 
\begin{equation}\label{E5}
\begin{aligned}
J_{n+\frac12}(x)&= \sqrt{\frac{2}{\pi x}} \left[\sin\left(x-\frac{\pi}{2}n\right)\sum_{k=0}^{\left\lfloor\frac{n}{2}\right\rfloor}\frac{(-1)^k (n+2k)!}{(2k)!(n-2k)!}\,(2x)^{-2k}\right.\\
&\qquad+\left.\cos\left(x-\frac{\pi}{2}n\right)\sum_{k=0}^{\left\lfloor\frac{n-1}{2}\right\rfloor}\frac{(-1)^k (n+2k+1)!}{(2k+1)!(n-2k-1)!}\,(2x)^{-(2k+1)}\right],
\end{aligned}
\end{equation}
appropriate for large $x$, may be used to generate an expansion for $f^{\rm{XZ}}(v, z)$ that is suitable for small $z$. We will be content to establish the leading-order term for small $z$, which derives from the leading-order term in the above representation for large $x$:
\begin{equation}\label{E6}
J_{n+\frac12}(x) \sim \sqrt{\frac{2}{\pi x}}\sin\left(x-\frac{\pi}{2}n\right), \qquad x\to \infty.
\end{equation}
Using Eq.~\eqref{E6} in Eq.~\eqref{E4} and keeping only the $m=0$ term, corresponding to the leading $x$-power in the integrand, we readily obtain
\begin{equation}\label{E7}
f^{\rm{XZ}}(v, z) \sim -8 v\,z\frac{\partial}{\partial z} \int_0^{\infty} dx\, x^3  \sin^2 x \frac{1}{e^{x\gamma z}-1}\sim -4 v \,z\frac{\partial}{\partial z}\, \Gamma(4) \zeta(4) (\gamma z)^{-4} = \frac{16\pi^4 v}{15\gamma^4 z^4}, \qquad z \to 0.
\end{equation}
It is interesting to note the appearance of the Planck-Einstein transformed temperature, $T_{\gamma}\equiv\frac{T}{\gamma}$, in this (high-temperature) limit. Note Eq.~\eqref{E7} captures not only the correct $ z $ dependence but also the velocity dependence of $ f^{\rm{XZ}} $ in the small $ z $ limit. The agreement of Eq.~\eqref{E7} with the numerical data for $ v=0.5 $ is illustrated in Fig.~\ref{FigE}.

Likewise, the infinite series representation
\begin{equation}\label{E8}
J_{\mu}(x) J_{\nu}(x)=\sum_{n=0}^{\infty}\frac{(-1)^n (\mu+\nu+n+1)_n}{n!\Gamma(\mu+n+1)\Gamma(\nu+n+1)}\left(\frac{x}{2}\right)^{\mu+\nu+2n},
\end{equation}
appropriate for small $x$, may be used to generate an expansion for $f^{\rm{XZ}}(v, z)$ that is suitable for large $z$. In this case, the leading $x$-power in the integrand in Eq.~\eqref{E4} is independent of $m$, so all terms must be included, resulting in 
\begin{align}
\label{E9}
f^{\rm{XZ}}(v, z) &\sim-\pi \sum_{m=0}^{\infty} \frac{v^{2m+1}}{m!} \frac{2^{-(3+2m)}}{\Gamma\left(\frac72\right)\Gamma\left(m+\frac72\right)} z^{2m+1}\frac{\partial^{2m+1}}{\partial z^{2m+1}}\int_0^{\infty}dx\, x^9 \frac{1}{e^{x\gamma z}-1}\nonumber\\
&=-\pi \sum_{m=0}^{\infty} \frac{v^{2m+1}}{m!} \frac{2^{-(3+2m)}}{\Gamma\left(\frac72\right)\Gamma\left(m+\frac72\right)} z^{2m+1}\frac{\partial^{2m+1}}{\partial z^{2m+1}} \,\Gamma(10) \zeta(10) (\gamma z)^{-10}\nonumber\\
&=\pi\sum_{m=0}^{\infty}\frac{v^{2m+1}}{m!}\frac{2^{-(3+2m)}\,(2m+10)! \,\zeta(10)}{\Gamma\left(\frac72\right)\Gamma\left(m+\frac72\right)} \frac{1}{(\gamma z)^{10}}\nonumber\\
&=\frac{2^8\zeta(10) \,v}{15\,\gamma^{10}z^{10}}\sum_{m=0}^{\infty}v^{2m}(m+1)(m+2)(m+3)(m+4)(m+5)(2m+7)(2m+9)\nonumber\\
&=\frac{2^{11} 3\,\zeta(10)}{z^{10}} \,\gamma^6 v \,(21+30v^2+5v^4), \qquad z\to \infty,
\end{align}
where we have used the identity
\begin{equation}\label{E10}
\gamma^{2n}=\frac{1}{(n-1)!}\frac{d^{n-1}}{d (v^2)^{n-1}}\frac{1}{1-v^2}=\frac{1}{(n-1)!}\sum_{m=0}^{\infty}v^{2m}(m+1)(m+2)\cdots (m+n-1).
\end{equation}
The result obtained in Eq.~\eqref{E9} is precisely that found in Eq.~\eqref{key}.

%references
\bibliography{qfcite}

\end{document}